\documentclass[sigconf]{acmart}

\usepackage{epsfig}
\usepackage{bbm}
\usepackage{url}
\usepackage{xcolor}
\usepackage{pgfplots}
\pgfplotsset{compat=1.14}

\newcommand\Warning{\makebox[1.4em][c]{\makebox[0pt][c]{\raisebox{.1em}{\small!}}\makebox[0pt][c]{\color{black}\Large$\bigtriangleup$}}}

\newcommand{\ignore}[1]{}
\newcommand{\authnote}[3]{\textcolor{#2}{{\sf (#1's Note: {\sl{#3}})}}}

\newcommand{\jenote}{\authnote{Jeremiah}{blue}}

\author{Ben Harsha}
\affiliation{%
	\institution{Purdue University}
	\city{West Lafayette}
	\state{Indiana}
}
\email{bharsha@purdue.edu}

\author{Robert Morton}
\affiliation{%
	\institution{Purdue University}
	\city{West Lafayette}
	\state{Indiana}
}
\email{rob.morton@me.com}

\author{Jeremiah Blocki}
\affiliation{%
	\institution{Purdue University}
	\city{West Lafayette}
	\state{Indiana}
}
\email{jblocki@purdue.edu}

\author{John Springer}
\affiliation{%
	\institution{Purdue University}
	\city{West Lafayette}
	\state{Indiana}
}
\email{jaspring@purdue.edu}

\author{Melissa Dark}
\affiliation{%
	\institution{Purdue University}
	\city{West Lafayette}
	\state{Indiana}
}
\email{dark@purdue.edu}

\begin{document}

\date{}

\title{Bicycle Attacks Considered Harmful: Quantifying the Damage of Widespread Password Length Leakage}

\begin{abstract}
We examine the issue of password length leakage via encrypted traffic i.e., bicycle attacks. We aim to quantify both the prevalence of password length leakage bugs as well as the potential harm to users. In an observational study, we find that {\em most} of the Alexa top 100 rates sites are vulnerable to bicycle attacks meaning that an eavesdropping attacker can infer the {\em exact} length of a password based on the length the encrypted packet containing the password. We discuss several ways in which an eavesdropping attacker could link this password length with a particular user account e.g., a targeted campaign against a smaller group of users or via DNS hijacking for larger scale campaigns. We next use a decision-theoretic model to quantify the extent to which password length leakage might help an attacker to crack user passwords. In our analysis, we consider three different levels of password attackers: hacker, criminal and nation-state. In all cases, we find that such an attacker who knows the length of each user password gains a significant advantage over one without knowing the password length. As part of this analysis, we also release a new differentially private password frequency dataset from the 2016 LinkedIn breach using a differentially private algorithm of Blocki et al. (NDSS 2016) to protect user accounts. The differentially private dataset contains frequency lists split by length which allows us to simulate an online attacker's Bayesian updates after obtaining the length of a password. The LinkedIn frequency corpus is based on over 170 million passwords making it the largest frequency corpus publicly available to password researchers. While the defense against bicycle attacks is straightforward (i.e., ensure that passwords are always padded before encryption), we discuss several practical challenges organizations may face when attempting to patch this vulnerability. We advocate for a new W3C standard on how password fields are handled which would effectively eliminate most instances of password length leakage.

\end{abstract}


\maketitle

\section{Introduction} \label{sec:Intro}
In any efficient encryption scheme there is necessarily some relationship between plaintext length and ciphertext length e.g., consider encrypting a 2MB jpeg image vs encrypting a 2GB mp4 movie. Vincent Guido~\cite{bicycleAttack} observed that (unpadded) SSL traffic can leak information about password lengths, and introduced the name ``bicycle attack'' in reference to the fact that a gift-wrapped bicycle still looks like a bicycle. Thus, whenever plaintext-length might be viewed as a sensitive attribute it is recommended that an application developer should pad the plaintext message before encryption~\cite{TLS12}. For example, the RFC for TLS 1.2 includes the following caveat:
	
	\begin{quote}Note in particular that type and length of a record are not protected by encryption.  If this information is itself sensitive, application designers may wish to take steps (padding, cover traffic) to minimize information leakage.  \end{quote}

Authenticated Encryption with Associated Data (AEAD) ciphers simultaneously guarantees both message integrity and confidentiality.  A recent longitudinal study of TLS Deployment found a dramatic rise in the percentage of TLS connections using AEAD cipher such as AES128-GCM, AES256-GCM and ChaCha20-Poly1305 since 2013~\cite{TLSDeployment} i.e., roughly $80\%$ of TLS connections used either AES128-GCM or AES256-GCM in April, 2018. In all three AEAD schemes there is a one to one relationship between the length of some ciphertext and the length of the original plain text message. This 1-1 relationship can be viewed as a feature of the cipher as it allows for significantly shorter ciphertexts i.e., a $2$ byte message would not need to be padded to a $16$-byte (AES128-GCM) or $32$-byte message (AES256-GCM) before encryption. However, the 1-1 relationship means that an eavesdropping attacker to infer the {\em exact} length of each transmitted message from the intercepted ciphertext. The responsibility of identifying cases where the length of a plaintext message is potentially sensitive and ensuring that such messages are appropriately padded is left to the application developer.\footnote{For example, the RFC for TLS 1.3~\cite{TLS13} explicitly says ``Selecting a padding policy that suggests when and how much to pad is  a complex topic and is beyond the scope of this specification.''}

Arguably, passwords provide a clear example where the message length is a sensitive attribute. Broadly speaking there are two categories of password guessing attacks: online attacks and offline attacks. In both attacks, the adversary attempts to gain access to some user's account by repeatedly guessing different possible passwords from a password cracking dictionary. While offline attacks are more dangerous they require the attacker to first hack into an authentication server to steal the password hash whereas online attacks can be mounted at any time. Even though an online attacker is rate-limited these attacks can be surprisingly effective, especially when the attacker has targeted background information about each user e.g., DOB, address, phone number~\cite{CCS:WZWYH16}. One defense is to rate limit the attacker by locking the account after several incorrect guesses within a fixed time window~\cite{apple_lockout} or by requiring the user to solve CAPTCHA challenges~\cite{kirk_2014} in between consecutive incorrect guesses. However, a stricter policy is more likely to inconvenience honest users who might forget/mistype their password. Furthermore, a password spraying attacker will attempt to overcome these restrictions by cycling through {\em many} different accounts that the attacker would like to crack e.g., the attacker might try $3$ guesses per day per user account while still attempting submitting thousands/millions of password guesses per day against a different user account. An online/offline password attacker who knows the length of a user's password can potentially obtain a significant advantage in eliminating guesses from the cracking dictionary of the wrong length.

In this paper we aim to answer two key research questions: 1) How many web sites are potentially vulnerable to bicycle attacks? 2) How does password length leakage impact the fraction of passwords that an attacker would crack in an online/offline attack?

\section{Contributions}

\paragraph{Observational Study} We conducted an observational study of AES-GCM traffic to see if application designers took appropriate steps (padding) to prevent password length leakage \footnote{There are of course several other ciphers with a 1-1 relationship between message length and ciphertext length. We chose to focus on AES-GCM as the cipher already accounts for over $80\%$ of TLS connections~\cite{TLSDeployment} and this number is likely to increase as the Internet continues to transition to TLS version 1.3.}  In our observational study we find that there is a widespread failure to pad passwords before encryption e.g., {\em at least} 84 of the Alexa Top 100 pages are vulnerable to bicycle attacks. As a concrete example, we demonstrate how an eavesdropping attacker can infer the exact lengths of Gmail passwords and Chase Bank passwords by monitoring encrypted traffic between the user and the authentication server. 

One challenge an eavesdropping attacker might face in exploiting this vulnerability is {\em linking}  specific usernames to IP addresses. For example, an eavesdropping attacker can easily extract the length of a password from the encrypted packets exchanged between a client (e.g., user) and server (e.g., Gmail/Chase Bank), but this information is less useful to an attacker if the packet (from which password length can be inferred) cannot be linked to a specific user account. A targeted eavesdropping attacker might choose to target specific users (e.g., by sniffing traffic on the home WiFi network), but for a password spraying attacker, this approach will not scale. If an eavesdropping attacker can exploit some other vulnerability to link usernames with client IP addresses in bulk then a password spraying attacker will also be able to infer the length of each user's password. We discuss how a sophisticated attacker could use DNS hijacking to extract usernames from unencrypted traffic and then associate those user names with client IP addresses in bulk. 


\paragraph{Quantifying the Damage of Bicycle Attacks} Motivated by the observation that many high profile web sites are vulnerable to bicycle attacks we aim to quantify the severity of this vulnerability by estimating how many additional user passwords an adversary might successfully guess before/after obtaining the lengths of each password. To answer this question we adapt a decision-theoretic model of a password cracking adversary of Blocki et al.~\cite{blockiCASH2016,SP:BloHarZho18} which was used to analyze the behavior of a rational offline password cracking adversary. A rational password cracking adversary will continue attempting to crack the user's password until the marginal cost of one more password guess ($k$) exceeds the marginal reward i.e., the value of the cracked password $(v)$ times the probability $(p_i)$ the next password guess is correct. In the context of an offline attack, the marginal guessing cost ($k$) is given by the cost of computing the password hash function, while in the context of an online attack the cost $k$ might be dominated by the cost of solving a CAPTCHA challenge. 

Given a probability distribution $p_1 \geq p_2 \ldots $ over user-selected passwords (here $p_i$ is the probability of the i'th most popular password) as well as the cost/value the parameters $v$ and $k$ the model allows us to predict how many passwords a rational attacker would crack. Of course, if the attacker learns the length $\ell$ of the user's password then we obtain a different prediction based on the updated probability distribution $p_1^\ell \geq p_2^\ell,\ldots$. Here, $p_i^\ell$ is the probability of the i'th most popular password of length $\ell$. Our model also allows us to estimate the attacker's expected utility (reward minus guessing cost) with and without knowledge of the password length. 

\paragraph{Differentially Private LinkedIn Password Frequency Corpus} In order to run the analysis we need to obtain empirical estimates of  $p_1 \geq p_2 \ldots $ and of  $p_1^\ell \geq p_2^\ell,\ldots$ for each length $\ell$. The RockYou dataset (32.6 million passwords) allows us to estimate these values, but the passwords in the dataset might have lower entropy since the account is arguably {\em low-value}. The larger differentially private Yahoo! frequency corpus (70 million passwords)~\cite{NDSS:BloDatBon16} is not split by length and as such will not allow us to estimate how many passwords the attacker will crack when the length $\ell$ is known. We address this problem by using a differentially private algorithm of Blocki et al.~\cite{NDSS:BloDatBon16} to publish a frequency corpus based on 174+ million cracked password hashes from the LinkedIn breach~\cite{linkedin_breach}. The frequency corpus contains several frequency lists $f_1 \geq f_2 \geq \ldots $ and $f_1^\ell \geq f_2^\ell \geq \ldots$ for each length $\ell$ where $f_i$ (resp. $f_i^\ell$) is the  number of user who selected the i'th most popular password of any length (resp. most popular password of length $\ell$). The frequencies are perturbed slightly to satisfy the mathematically rigorous notion of differential privacy~\cite{ICALP:Dwork06,TCC:DMNS06}. We use the LinkedIn frequency to obtain empirical estimates for the password distribution e.g., $p_i^\ell = f_i/N_\ell$ where $N_\ell$ is the total number of passwords of length $\ell$. This allows us to estimate {\em both} the number of additional user accounts that would be vulnerable to cracking when password lengths are leaked as well as the {\em utility gain} from to password length leakage. We believe that the frequency corpus will have tremendous independent value for password security researchers as it is more than twice as large as the Yahoo! corpus. 

\paragraph{Our Analysis}We consider three categories of attackers with different value to cost ratios $(v/k)$: a hacker $(v/k) \in \{10^2,10^3\}$, a criminal $(v/k) \in \{10^4,10^5\}$ and a nation-state $(v/k) \in \{10^6,10^7\}$. In all cases we find that password length leakage {\em significantly} increases the fraction of accounts vulnerable to an online attack. For example, our model predicts that a hacker with value to cost $v/k = 10^2$ will crack $2.8\%$ of passwords without knowledge of the password length, but this number jumps to $5.8\%$ when the length of a password is leaked --- an increase of $3\%$! For a nation-state attacker with a higher ratio $v/k=10^6$ we find that an attacker who knows the length of user passwords will crack an additional $19.5\%$ of user passwords! Similar results hold for the RockYou distribution.   

We also analyze the attacker's utility gain when the password length is leaked, which gives us an estimate on the value of the leaked length information. For example, for a criminal attacker with value/cost ratio $(v/k) = 10^5$ we find that the attacker's utility increases by $7203 \times k$ where $k$ is the cost of each password guess and this gain increases to $1.3 \times 10^5 \times k$ for a nation-state with $(v/k) = 10^6$. Thus, if we estimate that $k\approx \$0.001$ (e.g., the cost to pay a human user to solve a CAPTCHA) the criminal (resp. nation state) would choose to exploit the bug if the amortized cost of obtaining each target user's password length was  {\em at most} $\$7.2$ (resp. $\$130$). Given that the value of leaked length information is quite high we conclude that it would most likely be worthwhile for a criminal or nation-state attacker to obtain this information in an eavesdropping attack. 

\paragraph{Solutions} We demonstrate that leaking password lengths can significantly help a password cracking attacker. Nevertheless, the problem of password length leakage remains widespread. Given the severity of the bicycle attack vulnerability as indicated by our analysis above we argue that it is important to patch this bug quickly. Thus, we conclude by discussing potential solutions to the problem of widespread password length leakage. On an individual level, the solution is straightforward --- developers should always ensure that passwords are padded before they are encrypted. We provide a generic Javascript patch which could be implemented immediately by many web services. However, we also acknowledge patching a bug can be a slow process (e.g., see Heartbleed~\cite{HeartbleedPatch}), and that oftentimes patches introduce their own vulnerabilities. 

In general, we contend that it is inherently problematic to push the responsibility of padding plaintext messages entirely to the application developer whenever message length might be sensitive. We advocate for changes to W3C standards which could effectively eliminate bicycle attacks on passwords. In particular, we argue that web browsers should pad password fields by default before a web form is ever submitted. We also argue that other text input fields should take an optional parameter which specifies that the input length is sensitive.

\section{Related Work} \label{sec:Related}
Vincent Guido~\cite{bicycleAttack} previously introduced the name ``bicycle attacks'' in reference to his observation that unpadded SSL traffic can leak information about the length of a password. Our contribution is an observational study which demonstrates that such leakage is widespread across the internet, and an empirical analysis of leaked password datasets to quantify the advantage of a password cracking who knows the length of a user's password.

The phenomena of information leakage from encrypted traffic has a rich history~(e.g., \cite{SP:CWWZ10}, and Dyer et al. observed that, in practice, efficient countermeasures fail to prevent traffic analysis attacks~\cite{SP:DCRS12}. In past observational studies, attacks against encrypted VoIP communications were conducted by Wright et al. \cite{wright2007language,Wright2008a}.  Knowledge of the underlying variable bit encoding and length preserving ciphers were used to derive intelligence on the language spoken and known phrases.  Specifically, machine learning techniques were used to gain specific information about a ciphertext.  Further research by White et al. attempted to build upon this research to gain more generalized information about VoIP traffic \cite{White2011}. Goldberg et al. use the Gaussian-like unimodal distribution to retrieve latency between keystrokes and ultimately derive their variance \cite{Goldberg2009}. Separately, Frosch et al. used the total source-destination metric, edit-distance metric, and random metric to identify differences in encrypted traffic for health care websites \cite{frosch2016secure}. Unger and Goldberg discovered techniques to identify common user actions such as clicking a send button by analyzing the size of cipher messages on popular social networking sites like Facebook and Twitter \cite{Unger2015}. 

Other pertinent research includes Paul Kocher's investigation of timing attacks on public key crypto \cite{Kocher1996}, analysis of SSH keystrokes and timings by Song et al. \cite{song2001timing}, efforts by Fiore and Abadi in the exploration of symbolic techniques for examining cryptographic protocols \cite{fiore2001computing}, and later work in cryptographic protocols by Doghmi et al. \cite{doghmi2007searching}. Also relevant is work by Borisov et al. involving "off the record" protocols \cite{borisov2004off} and Chapman and Evans~\cite{Chapman2011} pertaining to blackbox detection of side-channel vulnerabilities. Additionally, vulnerabilities in mobile apps have been the foci of several researchers including \cite{Chen2015, Unger2015b, conti2015can,garman2016dancing}. Information leakage has also been a significant concern in Searchable Encryption as discussed by \cite{Cash2015} among others.

AlFardan et al.~\cite{USENIX:ABPPS13} showed how an eavesdropping attacker could recover cookies via Bayesian analysis exploited weaknesses in the RC4 cipher, though the eavesdropping attacker must intercept $2^{34}$ encryptions of the cookie to succeed. Garman et al.~\cite{USENIX:GarPatMer15} later improved these attacks and showed how an attacker can recover passwords from $2^{26}$ encryptions with RC4. Both attacks exploit weaknesses in the underlying RC4 cipher. While bicycle attacks only reveal the length of a password an eavesdropping attacker only needs to intercept {\em one} encryption of a password to infer the length. The attack does not rely on any underlying weakness of the underlying cipher (e.g., AES), but instead relies on the assumption that passwords are not being automatically padded before encryption.
\ignore{
\jenote{We might consider getting rid of this next paragraph. The focus of the paper is not QUIC or AES-GCM specifically, but length leakage in general. }
As for QUIC, past research on this protocol suggested that it does not satisfy forward secrecy requirements and any latency gains can be reversed with bit flipping or replay attacks \cite{Lychev2015}. Additional research on QUIC by Carlucci et al.~\cite{carlucci2015http} explored the performance improvements obtained by QUIC for transporting HTTP traffic. Well-known is the inherent trade-off between speed and security and QUIC proves no different. Moreover, as previous researchers did not quantify the loss in security, it is difficult for professionals to select the best cipher to make an informed decision about the trade-offs between speed and security. Concerns about AES-GCM are also well known including those discussed in \cite{EPRINT:McGVie04}. }

In addition to the previously mentioned research, the human propensity to select low-entropy passwords has been consistently documented over several decades~\cite{morris1979password,SP:Bonneau12}. Empirical studies have shown that complex password policies requiring users to use capital letters, numbers and special symbols in their passwords largely fail to produce stronger passwords \cite{Komanduri2011,Shay2014}, and complex policies adversely impact usability~\cite{Inglesant2010,Florencio2014lisa,Adams1999}. Wang et al.~\cite{CCS:WZWYH16} recently found that online attacks can be unexpectedly dangerous especially if an attacker has background knowledge (e.g., DOB, name, address) about the target user. 

Previous studies have explored the relationship between password {\em length} and password {\em strength} (e.g., ~\cite{ur2012does,SP:KKMSVB12} . By contrast, we are interested in studying how much damage occurs when this length is {\em leaked} to an attacker. To the best of our knowledge, this is the first paper to rigorously quantify the {\em damage} due to password length leakage.

\section{Prevalence of vulnerability}
A significant question to answer is just how prevalent password length leakage is in practice, and whether or not it is prevalent enough to be of concern. Our team first noted the issue during an observational study focusing on samples of network traffic captured from a VPN. During this observational study, it was noted that there was a 1-1 correspondence between the length of a plaintext message being sent and the ciphertext length. This was (in part) due to the use of the AES-GCM cipher, where the length of a ciphertext is in 1:1 correspondence with the length of the underlying plaintext. As we have remarked previously this relationship is an intentional feature of the AES-GCM cipher as it allows for shorter ciphertexts and it is the developer's responsibility to pad sensitive data (e.g., passwords) when length leakage is a concern i.e., see \cite{TLS12}. 

\paragraph{Initial Experiment: CiscoASA 5506} In our first experiment collected and analyzed network traffic produced from the Cisco ASA 5506 appliance, one of the most popular VPN devices used for connecting organizations around the world. The Cisco ASA VPN device was configured to use TLS version 1.2 to connect AnyConnect clients for remote access to the internal network. During collection, it was observed that the Cisco ASA 5506 switched from an encrypted TLS connection and used the Quick UDP Internet Connection (QUIC) protocol developed by Google to improve performance (the use of QUIC was absent in the Cisco documentation).

During this observational study it was noted that there was a 1-1 correspondence between the length of a plaintext message being sent and the ciphertext length. This was (in part) due to the use of the AES-GCM cipher in QUIC. In particular, the Cisco ASA VPN device was not padding plaintext messages which allowed us to easily identify specific events in an encrypted IRC chat session i.e., client connection/disconnection and to derive the length of individual IRC chat messages and discriminate between web traffic and ftp traffic. One can easily imagine a setting in which such leakage over a VPN network might be highly sensitive  e.g. linking two political dissidents who are chatting on IRC. 


\paragraph{Experiment 2: Password Length Leakage in Gmail} Motivated by the first observational study we sought to determine whether any prominent web pages might be vulnerable to bicycle attacks. In this study, we used the Chrome web browser, because, by default, QUIC is enabled with Chrome. We employed a collection platform using WireShark and allowed simultaneous collection on the internal and external interfaces to easily compare the encrypted and unencrypted traffic. We then used our automated process to login to Google Mail three times increasing the password length by one character every ten seconds. We had a ten-second delay between passwords entered as resulted in a pronounced signature when reviewing QUIC network traffic in WireShark. Google Mail uses reCAPTCHA to rate limit attackers after a number of incorrect login attempts. The ten-second delay allows us to manually solve each CAPTCHA challenge within the ten second time delay built into our observational study. We found that passwords were {\em not} being padded before encryption as there was a trivial 1:1 correspondence between password length and ciphertext length --- see figure \ref{fig:packetlength}. 

We then conducted a similar observational study for the banking website for JP Morgan Chase using Apple's Safari web browser to observe TLS version 1.2 traffic. In all of our experiments, the AES-GCM cipher was selected when we connected to JP Morgan Chase. The results were similar to Gmail except that there was a 1:2 correspondence between the number of characters in the password and the length of the ciphertext --- due to the use of Unicode encoding instead of ASCII.

\paragraph{Alexa Top 100 Web Sites.} Having found several prominent instances where password lengths were being leaked we set out to do a more comprehensive review of the Alexa Top 100 web sites~\cite{alexatop100}. The results were concerning. Gmail and JP Morgan Chase were not isolated examples. Instead, we found that {\em at least} 84 out of the top 100 Alexa web pages at the time were vulnerable to bicycle attacks as they failed to properly pad password fields when using the AES-GCM cipher. In most instances, there was a direct 1:1 relationship between password length and packet size (ASCII), but even in other instances (e.g., UNICODE) we are able to infer the exact password length from the ciphertext. The full details of this observational study including equipment, setup, and details about ciphertext / plaintext size ratios are located in the Appendix.

\paragraph{Exploiting the Vulnerability.} 
The ease of conducting such an attack makes it particularly concerning. It is quite possible to obtain ciphertext length for particular users by observing wireless network traffic in public places. Even more concerning is that we find that 84 of the Alexa top 100 sites leak plaintext length for password fields. Given that these sites comprise a significant portion of a typical user's online activity this attack becomes even easier to carry out. All an adversary needs to do it sit in a public area and collect wireless traffic. They can then identify which site a user is trying to access, look up the plaintext password length, and then use that knowledge to assist them in other attacks e.g. online password attacks. 

There may be additional difficulties in running an attack like this e.g. without prior knowledge it may be difficult to identify which usernames are associated with which password lengths. These can still be overcome in the context of more targeted online attacks. As an example, an adversary targeting someone specific may be able to send some traffic (like an email) to a user and see if they can spot them receiving the message. They may also be listening in at a more specific location (such as a home) where they already know the usernames of those inside. Because the username and password length combination must be associated to make any attacks feasible we will be primarily focused on targeted online password guessing attacks, rather than more general attacks such as an offline password attack following a password breach (even though length data is also useful to adversaries in these situations).

\textbf{Disclosure.} This vulnerability was reported to the United States Computer Emergency Response Team (US-CERT), Cisco, Google, Apple, Microsoft, and JP Morgan Chase bank. Apple will use this information to inform decisions about security transport protocols in future products. Google is working with the Internet Engineering Task Force (IETF) to look at ways to stop information leakage through the HTTP/2 standards forum and have notified the companies supporting the development of major web browsers.
\label{sec:observation_summary}
\section{Exploiting Length Leakage} \label{sec:exploitlengthleakage}

	In this section, we describe how a password attacker might exploit bicycle attacks to gain an advantage. To fully exploit the vulnerability an attacker needs to be able to eavesdrop on encrypted TLS traffic, identify which packet(s) contain encrypted passwords and link the source IP addresses in a packet header with a particular username. We use Google Mail to illustrate how an eavesdropping attacker can easily identify which packet contains the encrypted password. This allows the attacker to infer that some Google Mail user with a given source IP address has a password of length $\ell$. 
	
In Section \ref{subsec:attacksetup} we describe the attack setup under the assumption that the attacker has a target Google Mail user in mind and is able to link the source IP address with the target user name e.g., a targetted attacker might intentionally sniff traffic on a public WiFi network at a small coffee shop when the target user arrives.  In Section \ref{subsec:exploit} we describe how a password attacker will proceed once he learns the length of the target user's Google Mail password i.e., by eliminating passwords from a cracking dictionary with the wrong length. In Section \ref{subsec:exploit} we discuss how a sophisticated attacker might exploit the vulnerability on a more wide spread scale by using DNS Hijacking to pair username with password lengths at scale e.g., as part of a password spraying campaign.

\subsection{Setting up the Attack} \label{subsec:attacksetup}
{\bf Attacker Capabilities.} We first highlight the capabilities an attacker must have to infer password lengths from encrypted Google Mail traffic by exploiting padding failures. First, an adversary needs access to a transit point between the client and server to collect network traffic for analysis.  Furthermore, the adversary must be capable of determining the password length using signature-based detection from different transport security protocols: TLS version 1.2, 1.3, and QUIC. Because these signatures could be computed offline this is not a limiting assumption. 

We setup our experiment by register for a dummy Gmail account with a randomly chosen password, and then simulate a targeted eavesdropping adversary by logging into the account while sniffing network traffic on WireShark. A video demo of the attack is available at \url{https://youtu.be/TRkIymYJmzY}. 

\textbf{Step 1} - We start with a dictionary of common user passwords and choose a password at random from the dictionary subject to the constraint that the password is consistent with Gmail requirements, e.g., each password is between 8 and 60 characters.

\textbf{Step 2} - We sign-up for a dummy Gmail account using the password selected in step 1.  This simulates in a safe manner a user using one of the common passwords in the real-world.

\textbf{Step 3} - We sniff the network with WireShark and sign-in with our newly created Gmail dummy account with the password used in Step 2. We developed a signature that looks for the first QUIC packet under 300 bytes i.e., the packet which contains the encrypted password. The signature then determines the password length $\ell$ by subtracting any overhead.  After subtracting any overhead, we identify that for every increase in 1 character in the password length the encrypted traffic increased by exactly 1 byte or has a 1:1 ratio.

\subsection{Running the Attack} \label{subsec:exploit}
After the attacker learns the length $\ell$ of a user's password, the attacker selects a dictionary containing the $B$ most popular length $\ell$ passwords. The attacker can now run an automated script to crack the user's password. This automated attack is possible whether or not the attacker knows the length $\ell$ of the user's password. However, an attacker who knows the length $\ell$ can potentially crack the password much faster.

The particular way in which the attacker will exploit this leaked information will depend on the particular lockout mechanism adopted by the server. GMail uses CAPTCHAs~\cite{EC:vBHL03} to rate limit an online attacker i.e., by requiring the attacker to solve a CAPTCHA challenge\footnote{A CAPTCHA is a puzzle that should be easy for a person to solve, but infeasible for a computer to solve using state of the art techniques. In a text CAPTCHA the user might be challenged with an image consisting of moderately distorted numbers and letters along with the instructions ``You must type those letters manually to prove you are not a robot.'' Pinkas and Sander~\cite{CCS:PinSan02} proposed to mitigate the risk of online guessing is to require the client to provide a proof-of-work (e.g., the solution to a CAPTCHA puzzle) along with each login attempt~\cite{CCS:PinSan02}. There is a delicate balance between ensuring that most human users can solve the puzzles consistently~\cite{SP:BBFNJ10,CCS:BurMarMit11}, and ensuring that a computer cannot solve the challenge. Advances in machine learning and neural networks ~\cite{NDSS:GYCZLT16,CCS:YTFZFX18} have made it harder to strike this balance. } after one or more incorrect login attempts. In all of our experiments, we never encountered a scenario where we were locked out even when running a script that continuously attempts. This solution has the advantage that it is not susceptible to denial-of-service attacks~\cite{wolverton2002hackers} wherein a malicious party attempts to lock an honest user by repeatedly attempting to login with incorrect passwords. On the negative, a determined attacker may continue trying to crack the password for as long as s/he is willing to continue solving CAPTCHA puzzles (or paying other humans to solve them). 

\textbf{Step 2} - We have an automated bot that iterates through each password in the (pruned) dictionary and repeats the following steps: 

\textbf{Step 2.A} - Our script checks the page to see if there is a CAPTCHA challenge. When CAPTCHA occurs, we use an online CAPTCHA solving service to pay real humans to solve the CAPTCHA challenge and copy the solution into the web form. In our demo, we used the Rumola service to bypass Google's reCAPTCHA service.

\textbf{Step 2.B} - Our script enters the current password guess in the password field and we submit the web form to check for a successful login. The script repeats Step 2 until the login is successful. We remark that in all of our experiments we never encountered a scenario where our attack script was locked out.

{\bf Other Defenses Against Online Attacks.} There are several different ways that an organization might defend against online attacks on user passwords. One defense is to lock an account after several consecutive incorrect login attempts. Lockouts can either be hard or soft. A hard lockout requires the user to authenticate via an additional factor, e.g., a security code sent via e-mail/text message, a phone call and/or answering a few security questions. A soft lockout expires after a fixed period of time, e.g., one hour or day. Stricter lockout policies are associated with usability problems~\cite{brostoff2003ten}, and are more vulnerable to denial-of-service attacks (e.g., an attacker can lock a user out of his account by repeatedly attempting incorrect logins)~\cite{wolverton2002hackers}. 

Pinkas and Sander~\cite{CCS:PinSan02} observed that an untargeted attacker with many usernames may avoid lockouts by alternating guesses between different accounts, and a determined attacker targeting a particular individual facing a soft lockout policy can avoid a lockout by rate limiting guessing attempts. We remark that if successful login attempts are visible to an attacker (e.g., via posts to social media) then even an attacker facing a hard lockout policy can avoid lockouts by only attempting to login after each successful user login. We stress that in either case (soft/hard lockout or CAPTCHAs) the attacker will benefit if s/he knows the length of the user's password.




\subsection{Linking Usernames and Password Lengths} \label{subsec:link}
To run the attack in Section \ref{subsec:exploit} an attacker would need to associate a user name for a specific website domain (gmail.com) with a specific client IP address. In this section, we demonstrate a plausible scenario in which a sophisticated attacker could link usernames and passwords with Google Mail using DNS Hijacking. This is simply one plausible scenario in which an intruder can pair username and password length pairs at scale e.g., as part of a password spraying attack.   
	
We remark that user ids often follow a predictable pattern e.g., jsmith for John Smith or jdoe for Jane Doe. Thus, any vulnerability that allows an attacker to link IP addresses to a name (or user id) will often allow the attacker to directly link the IP address to a specific user id. Nation-state attackers might coerce an internet service provider (ISP) to reveal this information directly, and a criminal might be able to extract this information by infiltrating the ISP.	
	
{\bf DNS Hijacking.} An attacker would start by  creating a DNS redirector by hijacking a legitimate DNS server:

The DNS Redirector is an attacker operations box which responds to DNS requests.

Step 1: The attacker would compromise legitimate DNS servers through already stolen credentials~\cite{department_of_homeland_security}.

Step 2: The attacker logs into a proxy box used to conduct non-attributed browsing and as a jumpbox to other infrastructure.

Step 3: A DNS request for target[.]com is sent to OP1 (based on previously altered A Record or NS Record).

If the domain is part of target[.]com desired domain, OP1 responds with an attacker-controlled IP address, and the user is redirected to the attacker-controlled infrastructure.

Step 4:  The attacker would then install a signature rule (e.g., SNORT rule) on OP1 to harvest cookies in unencrypted HTTP traffic~\cite{EFF,SP:SivPolKer16} and record the associated client IP address with each individual cookie.  The attacker would then perform offline analysis to discover usernames within the harvested cookies.  The attacker would then be able to build a list of associated user names with their client IP addresses.  

Step 5: The attacker would intercept on OP1 all HTTPS traffic for target[.]com/login.  In this, the attacker would then analyze encrypted AES-GCM traffic to discover the password length.  The attacker would record the password length with the client IP address.

Step 6:  The attacker would perform offline analysis to associate the user name from Step 4 and the password length from Step 5.

\paragraph{Discussion}  A 2016 study from Columbia researchers conducted an IRB-approved 30-day observation of their public wireless campus network and were able to intercept cookies containing 282K accounts~\cite{SP:SivPolKer16}. At that time the researchers found that all top Alexa websites exposed usernames in unencrypted HTTP traffic. The researchers found functionality across multiple cookies with different scopes between encrypted HTTPS and unencrypted HTTP traffic is complicated creating opportunities for exposure of private information such as usernames over unencrypted HTTP traffic~\cite{SP:SivPolKer16}. Despite the impressive progress of the ``Encrypt Everywhere'' movement over the last 4 years a substantial fraction (i.e., at least 20\%) of global web traffic is still unencrypted~\cite{EFF}. 

\subsection{An Improved Online Attack} \label{subsec:exploit}
Suppose that our eavesdropper has inferred the length $\ell$ of an encrypted password and {\em linked} that length with a specific user account. Our password attacker can immediately prune his cracking dictionary to remove any password that does not have length $\ell$.

In our observational study we found that GMail uses CAPTCHAs~\cite{EC:vBHL03} to rate limit an online attacker i.e., by requiring the attacker to solve a CAPTCHA challenge after each incorrect login attempt. 

We wrote a simple proof of concept script which automates the process of an online attack against gmail. After the attacker learns the length $\ell$ of a user's password, the attacker prunes his cracking  dictionary to focus on the most popular passwords of length $\ell$\footnote{Of course the same automated online attack is feasible whether or not the attacker knows the length $\ell$ of the user's password. However, an attacker who knows the length $\ell$ can potentially crack the password much faster.}. Before submitting each password guess our script checks the page to see if there is a CAPTCHA challenge\footnote{A CAPTCHA is a puzzle that should be easy for a person to solve, but infeasible for a computer to solve using state of the art techniques. In a text CAPTCHA the user might be challenged with an image consisting of moderately distorted numbers and letters along with the instructions ``You must type those letters manually to prove you are not a robot.'' There is a delicate balance between usability and security i.e., ensuring that a human can solve the challenge easily while a computer program cannot. Advances in machine learning and neural networks ~\cite{NDSS:GYCZLT16,CCS:YTFZFX18} have made it harder to strike this balance. Several  empirical  studies have demonstrated that CAPTCHA challenges are often difficult for human users to solves \cite{SP:BBFNJ10,CCS:BurMarMit11}.  } . Whenever CAPTCHA occurs, the attacker can use an online CAPTCHA solving service to bypass Google's reCAPTCHA service by paying human users to solve each challenge --- we used the Rumola service in our demo. The script continues to submit guesses until authentication is successful or until the cracking dictionary is exhausted.

We remark that in our experiments, we never encountered a scenario where we were locked out when running our attack script. Pinkas and Sanders~\cite{CCS:PinSan02} suggested the use of proof of work puzzles (such as CAPTCHAs~\cite{EC:vBHL03}) to rate limit an online attacker. One of the advantages of this approach (in comparison to a soft/hard lockout policy) is that it is not  susceptible to denial-of-service attacks~\cite{wolverton2002hackers} e.g., a malicious party might attempt to lock an honest user's account by repeatedly submitting incorrect login attempts. The downside is that a sufficiently determined attacker might continue trying to crack the password by paying humans to solve the CAPTCHA challenges.

{\bf Other Defenses Against Online Attacks.} There are several different ways that an organization might defend against online attacks on user passwords. One defense is to lock an account after several consecutive incorrect login attempts. Lockouts can either be hard or soft. A hard lockout requires the user to authenticate via an additional factor, e.g., a security code sent via e-mail/text message, a phone call and/or answering a few security questions. A soft lockout expires after a fixed period of time, e.g., one hour or day. Stricter lockout policies are associated with usability problems~\cite{brostoff2003ten}, and are more vulnerable to denial-of-service attacks (e.g., an attacker can lock a user out of his account by repeatedly attempting incorrect logins)~\cite{wolverton2002hackers}. 

Pinkas and Sander~\cite{CCS:PinSan02} observed that an untargeted attacker with many usernames may avoid lockouts by alternating guesses between different accounts, and a determined attacker targeting a particular individual facing a soft lockout policy can avoid a lockout by rate limiting guessing attempts. We remark that if successful login attempts are visible to an attacker (e.g., via posts to social media) then even an attacker facing a hard lockout policy can avoid lockouts by only attempting to login after each successful user login. We stress that in either case (soft/hard lockout or CAPTCHAs) the attacker will benefit if s/he knows the length of the user's password.

\section{Attacker Model}
	We consider three types of attackers, their capabilities, and motivations to compromise an account.  This will help us create a threat model before gaining information leaked from GCM.   We then can compare any increases in capability or willingness to understand the severity of the information leakage.

\begin{itemize}
	\item{\textbf{Hacker}} The first attacker is a hacker and is capable of controlling local networks, either on the Sender or Recipient side of the connection.  The hackers are motivated by the challenge of compromising an account and does not look to directly profit from the hack.  The hacker's willingness to compromise an account waivers in the face of systems with multiple security controls or a monetary cost of more than 0-100 (USD), because the account's perceived value to hacker's ego does not exceed its cost of 100.

\item{\textbf{Criminal}} The second attacker is a criminal with the capability to control local networks and several organizational wide networks.  The criminal is financially motivated to only compromise accounts whereby the payoff is at least ten times the cost \cite{InternetCrimeComplaintCenter2015}.   In most cases, the criminal will spend between 100-1,000 (USD) to compromise an individual account, because its perceived value to financially profit does not exceed its cost of 1,000 (USD).

\item{\textbf{Nation-State}} The third attacker is a nation-state.  The nation-state can compromise local and organizational networks and also control large segments of the Internet such as Internet Service Providers.  The nation-state is primarily motivated by understanding threats to its citizens or interests and is only willing to conduct multi-year campaigns against accounts threatening its security or sovereignty.   In most cases, we assume the nation-state is willing to spend between 1000-10,000 (USD) to compromise an individual account because the perceived value in the information gained from the compromised account does not exceed its cost of 10,000.  While we do not know the targeted account's subjective worth, we assume a rational adversary is not willing to pay more than its perceived value. 

A nation-state (and perhaps a sophisticated criminal) might have the capability to eavesdrop on network traffic on a broader scale to obtain (username, password length) pairs at scale e.g., using DNS Hijacking attack as discussed in Section \ref{subsec:link}. Similarly, a nation-state (and perhaps a sophisticated criminal) might have the ability to coerce an internet service provider to reveal logs associating specific users with the different IP addresses they were assigned over time.

Password spray campaigns typically target single sign-on (SSO) and cloud-based applications using federated authentication.  For example - in February 2018, the United States Department of Justice indicted nine Iranian nationals a part of the Mabna Institute for computer intrusions using password spraying \cite{department_of_homeland_security}.  While nation-states might be intrinsically motivated to conduct persistent and long-term campaigns, would an extrinsically motivated attacker being as willing to use a password-spraying technique?  To help answer this question, we conduct an economic analysis.  If the information leaked doubles an attacker's capability or willingness to conduct the attack, then we consider the information leak to be severe, because our data suggests doubling the adversarial advantage begins to show a clear change in the related monetary, decision-making, and temporal characteristics.  We want to create a threat model whereby the attackers have a spectrum of capabilities and willingness to compromise targets of interest.  Examples of severe information leakage might include the compromise of the session key, allowing an attacker to understand the underlying plain text without decryption, or increasing the capability beyond their stated initial set of capabilities to bypass a security control.  

\end{itemize}

	 

 \label{sec:pwdcrack}

\section{LinkedIn Password Frequency Corpus}

We are releasing several differentially private frequency lists derived from $N=174+$ million cracked passwords from the LinkedIn breach in 2012. The protocol was reviewed by the ethics board at our institution. In this section, we describe the LinkedIn dataset, the published frequency corpus and the differentially private algorithm that was used to release it. The frequency corpus is available through the (currently anonymous) link \url{https://figshare.com/articles/linkedin_files_zip/7350287}. 

We remark that the Yahoo! frequency corpus~\cite{SP:Bonneau12,NDSS:BloDatBon16} is significantly smaller ($N\approx =70$ million) and is not suitable for our analysis in Section \ref{sec:SecPrivacy} because the password frequencies cannot be linked with the length of each password. We use the LinkedIn dataset in Section \ref{sec:SecPrivacy} to quantify the damages of password length leakage. We also anticipate that the LinkedIn dataset will be of independent research interest for password analysis (e.g., see \cite{SP:BloHarZho18}) as it is significantly larger than the Yahoo! frequency corpus.

\paragraph{Background on LinkedIn Password Breach:} In 2012 hackers were able to compromise LinkedIn's systems and stole over 177.5+ million unsalted SHA-1 password hashes as well as other confidential user data. LinkedIn responded to the leak by releasing a public statement saying that they had invalidated compromised users' passwords, requiring them to be changed and by encouraging users to adopt two-factor authentication (see \url{https://blog.linkedin.com/2016/05/18/protecting-our-members}). Because the passwords were stored in unsalted form and LinkedIn was not using key-stretching tools such as PBKDF2 or Argon2~\cite{biryukov2016argon2} at the time of the breach the passwords were significantly easier to crack. KoreLogic, a cybersecurity company, cracked over $98\%$ of these password hashes \url{https://blog.korelogic.com/blog/2016/05/19/linkedin_passwords_2016}, and agreed to allow us to generate an anonymized frequency corpus based on these cracked passwords using the differentially private.

\paragraph{Password Frequency List: } A password frequency list is a non-negative list of integers $f_1 \geq f_2 \geq \ldots \geq 0$ where $f_i$ denotes the number of users in a dataset who selected the $i$\textsuperscript{th} most popular password and $N= \sum f_i$ denotes the total number of users in the dataset. We also use $f_1^\ell \geq f_2^\ell \geq \ldots \geq 0$ to denote the frequency list for length $\ell$ passwords, and $N_\ell = \sum f_i^\ell$ to denote the total number of users who selected a password of length $\ell$. Here, $f_i^\ell$ denotes the number of users who selected $i$\textsuperscript{th} most popular password with length $\ell$.

As an example consider a toy scenario in which $12$ users create LinkedIn passwords. Suppose that $f_1= 5$ users select the password ``123456,'' $f_2=3$ users select the password ``password'' and $f_3 = 2$ users select the password ``abc123'' and $f_4 = 2$ users select the password ``letmein.'' In this case the overall frequency list would be $(f_1,f_2,f_3,f_4) = (5,3,2,2)$ with $N=12$, but the frequency list for length $\ell = 6$ passwords would be $(f_1^6,f_2^6)=(5,2)$ with $N_6 = 7$ total passwords of length $6$.

\paragraph{Security Risks:} We first discuss the potential risk of releasing password frequency lists without any noise. In the example above the attacker happened to know the passwords for eleven of these users. If only four of these eleven users selected the password ``123456'' then the attacker who obtains the exact frequency list $(f_1,f_2,f_3,f_4) = (5,3,2,2)$ above would be able to infer the password of the remaining user. While this scenario may seem a bit far-fetched there are many examples of supposedly anonymized datasets that were later de-anonymized when the attacker has some background knowledge about the data e.g., \cite{SP:NarShm09}.

\paragraph{Differential Privacy} We use the notion of differential privacy~\cite{dwork2006calibrating,mcsherry2007mechanism} to ensure that the password statistics we release will not harm individual users. Differential privacy provides a strong information-theoretic privacy guarantees to each individual in a dataset, and it has been an active area of research in the last decade (e.g., see \cite{FOCS:McSTal07}). Informally, a differentially private mechanism for releasing statistics about a dataset $D$ ensures that an adversary cannot use the output to make inferences about any individual in the dataset except inferences that the adversary would have been able to make without that individual's data. Formally, a mechanism $\mathcal{A}$ which takes as input a frequency list $f_1,f_2,\ldots$ with $f_i \geq f_{i+1}$ and outputs a noisy frequency list is $(\epsilon,\delta)$-differentially private if for all subsets $S$ of output frequency lists we have

\[\Pr\left[\mathcal{A}\left(f\right) \in S \right] \leq e^{\epsilon} \Pr\left[\mathcal{A}\left(f'\right) \in S \right] + \delta \ , \]

whenever $f$ and $f'$ are neighboring frequency lists which satisfy the condition $\sum_i |f_i - f_i'| \leq 1$.

Here we can think of $f$ as the original frequency list and $f'$ as denoting the frequency list after a particular user's password is removed. Differential Privacy cannot promise users that an attacker won't crack their password after we release the frequency list i.e., it is possible that the user picked a weak password that would have been cracked in any case. However, we can promise each user that the probability of this outcome wouldn't have changed much even if we excluded their data entirely.

\paragraph{Exponential Mechanism for Differentially Private Frequency Lists} We used a differentially private algorithm $ \mathcal{E}_{\epsilon,\delta}$ developed by Blocki et al.~\cite{NDSS:BloDatBon16} to release the LinkedIn frequency corpus. The mechanism is highly accurate and guarantees that (whp) L1 distortion is very small. In particular, with high probability it holds that 
\[ \sum_{i} \left| f_i - \tilde{f}_i\right| \leq O\left( \frac{\sqrt{N}+\ln\left(\frac{1}{\delta} \right)}{\epsilon} \right) \ , \] 
where $\tilde{f} \leftarrow \mathcal{E}_{\epsilon,\delta}(f)$ is the output frequency list and $f$ is the original frequency list. 

We used the algorithm to publish $\tilde{f}$ (the original frequency list) as well as $\tilde{f}^\ell$ (the frequency list for passwords of length $\ell$) for each password length value $\ell$. Each individual password can contribute to {\em exactly} two of these frequency lists. Thus, we used the privacy parameter $\epsilon' = 0.25$ to release each individual frequency list to give an aggregate privacy value of $\epsilon = 0.5$. In all cases, we set $\delta \leq 2^{-100}$ so that this term was negligibly small.

\ignore{

\subsection{Stats}

Th

Total Passwords: 174292189

Number of Different Passwords: 57431283

$\lambda_{1} = 0.00651741197650573$

$\lambda_{10} = 0.0128475751715988$

$\lambda_{100} = 0.027981357213891$

$\lambda_{1000} = 0.0713368686877873$

$\lambda_{10000} = 0.142351479675317$

$\lambda_{100000} = 0.240106927568639$

$\lambda_{1000000} = 0.376761445115498$

$\lambda_{10000000} = 0.57864975809407$

Length 6:

Total Passwords: 33172884

Number of Different Passwords: 5418646

Length 7

Total Passwords: 23626429

Number of Different Passwords: 6672366

Length 8

Total Passwords: 45655199

Number of Different Passwords: 14886616

Length 9

Total Passwords: 22664131

Number of Different Passwords: 8750881

Length 10

Total Passwords: 16761091

Number of Different Passwords: 7505262

Length 11

Total Passwords: 7838163

Number of Different Passwords: 3814033

}

\section{Security and Privacy Impact of Password Length Leakage} \label{sec:SecPrivacy}

In this section, we aim to quantify the damages of password length leakage. In our analysis, we suppose that an online password attacker attempts to crack the user's password by repeatedly attempting the most popular passwords from a dictionary. We assume that the authentication server uses secure CAPTCHAs to rate limit the attacker (e.g., Gmail authentication). Thus, an attacker must pay human workers to solve a CAPTCHA after each incorrect guess. If the attacker knows the length of the password in advance then the attacker can eliminate passwords from the dictionary. 

We aim to answer the following questions: (1) How many additional passwords will an online attacker crack when given the length of each password? (2) How much does password length leakage monetarily benefit the attacker? We stress that questions (1) and (2) actually ask very different questions. The answer to the first question tells us how many additional user accounts will be compromised if password lengths are revealed to a rational attacker. The answer to the second question allows us to predict whether or not the cost of eavesdropping (equipment, manpower) on network traffic outweighs the benefit of learning password lengths. To address these questions, we adapt a game-theoretic model introduced by Blocki and Datta~\cite{blockiCASH2016} to model an offline password attacker.

\subsubsection{Constrained Attacker} 
Before introducing our decision-theoretic model, we first consider a constrained online password attacker who either gives up or gets locked out after $B$ incorrect guesses. This will give us the chance to introduce key notation. 

{\noindent \bf Notation} Let $\mathcal{D}$ denote the distribution over user selected passwords $\mathcal{P}$ and we let $p_i=\Pr_{pwd \leftarrow \mathcal{D}}[pwd=pwd_i]$ denote the probability that a user selects the $i$'th most popular password $pwd_i \in \mathcal{P}$ e.g. ,$p_i \geq p_{i+1}$. We also $\mathcal{P}^\ell = \{x \in \mathcal{P}: |x|=\ell \} \subseteq \mathcal{P}$ denote the set of all passwords with length $\ell$ and \[p_i^\ell = \Pr_{pwd \leftarrow \mathcal{D}}\left[pwd=pwd_{i_\ell}~\vline ~pwd \in \mathcal{P}^\ell\right] \] where $i_\ell$ is the index of the $i$'th most popular password in the set $\mathcal{P}^\ell$. Observe that we have $p_1^\ell \geq p_2^\ell \geq \ldots$ for each $\ell \geq 1$. For notational convenience we also write $\Pr\left[ \mathcal{P}^\ell\right] = \Pr_{pwd \leftarrow \mathcal{D}}\left[pwd\in  \mathcal{P}^\ell\right] $.

\subsubsection{Experiment 1: Unknown Lengths with $B$ Guesses.} A user selects a random password $pwd \leftarrow \mathcal{D}$ and an attacker attempts to guess the password online. We assume an attacker who knows the distribution $\mathcal{D}$, but not the specific password and that the attacker either gives up or gets locked out after $B$ guesses. The attacker best strategy is to try the $B$ most likely guesses in the distribution $pwd_1,\ldots,pwd_B$. The attacker will succeed with probability $\lambda_B = \sum_{i =1}^B p_i$. Conditioning on the event that the user's password has length $\ell$ (i.e., $pwd \in \mathcal{P}^\ell$) and that the length $\ell$ is {\em unknown} to the attacker, the attacker will succeed with probability  \[ \lambda_{B,\ell} = \Pr_{pwd \leftarrow \mathcal{D}}\left[pwd \in \bigcup_{i=1}^B \{ pwd_i \}~\vline ~pwd \in \mathcal{P}^\ell\right] \ .\]

\subsubsection{Experiment 2: Known Lengths with $B$ Guesses} This experiment is exactly like experiment one except that after we sample $pwd \leftarrow \mathcal{D}$ the length $\left| pwd\right|$ of the password is revealed to the attacker. If the attacker knows the password length is $\ell$ (i.e., $pwd \in \mathcal{P}^\ell$), then the attacker succeeds with probability 
\[ \lambda_{B,\ell}^* = \sum_{i =1}^B p_i^\ell \ .\]
If the attacker attempts the $B$ most likely guesses before giving up then he will succeed with probability
\[ \lambda_{B}^* = \sum_{\ell} \Pr[ \mathcal{P}^\ell ] ~\lambda_{B,\ell}^*   \ . \] 

\paragraph{Analysis} Table \ref{tab:limitadv} compares the success rate of the attacker with $(\lambda^*_B)$ and without $(\lambda_B)$ knowledge of the passwords length for various guessing limits $B$. The results show that the attacker's success rate increases significantly when the password length is known, e.g., a criminal attempting $B=10^5$ guesses per user using the LinkedIn distribution would crack nearly $35\%$ of passwords with knowledge of password length compared to $24\%$ without this knowledge, or about $50\%$. Table \ref{tab:limitstats} compares the attacker's conditional success rate with $(\lambda^*_{B,\ell})$ and without $(\lambda_{B,\ell})$ knowledge of the passwords length conditioning on the event that the user's password has length $\ell$\footnote{Table \ref{tab:limitstats} only shows information from the Rockyou leak. This is because the calculation of $\lambda_{B,\ell}$ requires a list showing the exact order of passwords as well as their frequencies. It is not enough to know something like ``there are 5 passwords picked with frequency 100". While you could identify which lengths made up those 5, we do not have the specific order. It is possible to construct an estimated list by arbitrarily ordering those 5 passwords, but this would produce a noisy estimate. Thus we do not have accurate values of $\lambda_{B,\ell}$ for the LinkedIn data}.  Surprisingly, even for longer lengths such as $\ell=30$ an attacker still may have a reasonably high success rate with a smaller guessing limit $B$. For example, $\lambda_{B,\ell}^* \approx 4.5\%$ at just $B=10$ guesses against the Rockyou list when password length is a large as $\ell=30$.\footnote{Of the $1,052$ users in the RockYou dataset with length $\ell=30$ passwords eight users selected ``bebelicouz\_05\_mistme@yahoo.com'' and seven users selected ``111111111111111111111111111111.'' Another popular password of length $\ell=30$  was the url \url{http://www.rockyou.com/tos.php} --- selected by five users.} By contrast, when the attacker does not know the length we have $\lambda_{B,\ell} = 0$ when $\ell=30$ even if the attacker tries up to $B=288,046$ guesses! This is because there are $288,046$ passwords with length $\ell \neq 30$ that are more popular than the most popular password of length $30$. 

\paragraph{Limitations of the LinkedIn and Rockyou Datasets.} 
A general limitation of empirically defined password distributions is that they almost certainly overestimate the probability of passwords at the tail of the distributions, e.g., for {\em any} password $pwd_i$ that was observed once in the Linkedin dataset we estimate that $p_i$ is at most $p_i = 1/N \geq  5.73 \times 10^{-9}$. The Linkedin dataset contains $1.7 \times 10^8$ unique passwords and about $2.1 \times 10^7$ of these passwords are observed exactly one time. Unfortunately, there is no way to be confident about the true probability of an event that has only been observed once. Thus, for $B > 1.5 \times 10^7$ our estimate of $\lambda_B$ may be too high, and for $B > 3.1 \times 10^6$ our estimate for $\lambda_{B,6}^*$ may be too high (there are about $3.1 \times 10^6$ length six passwords that were observed more than once so our estimate of $p_i^6$ may be too high for $i > 3.1 \times 10^6$).  When $B \leq 10^6$ we believe the estimate for $\lambda_B^*$ is reasonable\footnote{For each $\ell \in [6,10]$ we have more than $1.6 \times 10^6$ ($2 \times 10^5$) passwords of that length that were observed multiple times. These lengths account for $\sum_{\ell=6}^{10} \Pr[\mathcal{P}^\ell] \approx 81.4\% (86.5)\%$ of all passwords in the Linkedin (Rockyou) dataset. If we include $\ell=5,11$ then we have  $\sum_{\ell=5}^{11} \Pr[\mathcal{P}^\ell] \geq 85.9\% (94.1\%)$ and for both lengths $\ell=5,11$ we have over $7.8 \times 10^6$ ($8.75 \times 10^4$) passwords of that length that were observed multiple times.}. This means that there is some uncertainty about our estimates of $\lambda_{B}^*$ for a nation-state attacker ($B\in [10^6,10^7]$). In our analysis, we use the symbol \Warning to indicate that it is affected by uncertainty about the tail of the password distribution. 

As part of our analysis, we used empirical data from the RockYou dataset~\cite{cubrilovic_2009}, released in 2009,  to define our password distribution. The dataset was released in 2009 by hackers and remains one of the largest available \textbf{plaintext} password datasets. One potential downside is that many RockYou users may have viewed their account as low-value. While Bonneau~\cite{bonneau2012science} found that account value did not appear to be correlated with password strength in his analysis of Yahoo! passwords, we cannot rule out the possibility that RockYou users were less motivated to pick strong passwords because the account had low-value. However, we remark that it is possible that a stronger password distribution would result in an even bigger advantage $\lambda^*_B - \lambda_B$ for an attacker who learns the password length since both $\lambda^*_B$ and $\lambda_B$ would decrease.

\begin{table}

\caption{Attacker Success Rate at Various Guessing Limits with and without knowledge of the password length (LinkedIn) \label{tab:limitadv}}

\resizebox{\columnwidth}{!}{%
\begin{tabular}{|l|c|c|c|}
\hline
\multicolumn{4}{|c|}{LinkedIn} \\
\hline
Adversary type	&	Hacker	& 	Criminal	& Nation-state \\ \hline

Guess limit $B$	&	\begin{tabular}{l|l} $10^2$	& $10^3$ \end{tabular}
			&	\begin{tabular}{l|l} $10^4$	& $10^5$ \end{tabular}
			&	\begin{tabular}{l|l} $10^6$	& $10^7$ \Warning \end{tabular} \\ \hline
$\lambda^*_{B}$	
			& \begin{tabular}{r|r} 0.058	& 0.119 \end{tabular}
			& \begin{tabular}{r|r} 0.214	& 0.352 \end{tabular}
			& \begin{tabular}{r|r} 0.571	& 0.973 \end{tabular} \\ \hline			
$\lambda^*_{B}-\lambda_B$	
			& \begin{tabular}{r|r} 0.030	& 0.048 \end{tabular}
			& \begin{tabular}{r|r} 0.072	& 0.112 \end{tabular}
			& \begin{tabular}{r|r} 0.195	& 0.394 \end{tabular} \\ \hline
			
$\lambda^*_{B}/\lambda_B$	
			& \begin{tabular}{r|r} 2.074	& 1.672 \end{tabular}
			& \begin{tabular}{r|r} 1.505	& 1.466 \end{tabular}
			& \begin{tabular}{r|r} 1.517	& 1.682 \end{tabular} \\ \hline

\multicolumn{4}{|c|}{Rockyou} \\
\hline
Adversary type	&	Hacker	& 	Criminal	& Nation-state \Warning \\ \hline

Guess limit $B$	&	\begin{tabular}{l|l} $10^2$	& $10^3$ \end{tabular}
			&	\begin{tabular}{l|l} $10^4$	& $10^5$ \end{tabular}
			&	\begin{tabular}{l|l} $10^6$	& $10^7$ \end{tabular} \\ \hline
$\lambda^*_{B}$	
			& \begin{tabular}{r|r} 0.089	& 0.192 \end{tabular}
			& \begin{tabular}{r|r} 0.330	& 0.519 \end{tabular}
			& \begin{tabular}{r|r} 0.796	& 1.000 \end{tabular} \\ \hline			
$\lambda^*_{B}-\lambda_B$	
			& \begin{tabular}{r|r} 0.043	& 0.080 \end{tabular}
			& \begin{tabular}{r|r} 0.107	& 0.153 \end{tabular}
			& \begin{tabular}{r|r} 0.255	& 0.133 \end{tabular} \\ \hline
			
$\lambda^*_{B}/\lambda_B$	
			& \begin{tabular}{r|r} 1.938	& 1.712 \end{tabular}
			& \begin{tabular}{r|r} 1.479	& 1.418 \end{tabular}
			& \begin{tabular}{r|r} 1.471	& 1.154 \end{tabular} \\ \hline

\end{tabular}	
}	

\end{table}
\begin{table}
\caption{Attacker Conditional Success Rate at Various Guessing Limits with and without knowledge of the password length}\label{tab:limitstats}

\centering
\resizebox{\columnwidth}{!}{%
\begin{tabular}{|l|c|c|c|c|}
\hline
Lengths		&	$\lambda_{B,\ell}^*$	&	$\lambda^*_{B,\ell} - \lambda_{B,\ell}$ 	&	$\lambda^*_{B,\ell}  / \lambda_{B,\ell}$ \\ \hline

limit		&	

	\begin{tabular}{r|r|r}
		$10^2$	&	$10^4$	&	$10^6$\Warning
	\end{tabular}
				&
	\begin{tabular}{c|c|c} 
		$10^2$	&	$10^4$	&	$10^6$\Warning
	\end{tabular}
				&
	\begin{tabular}{c|c|c} 
		$10^2$	&	$10^4$	&	$10^6$\Warning
	\end{tabular}
	\\ \hline

5		&	

	\begin{tabular}{r|r|r} 
		0.228	&	0.670	&	1.000 
	\end{tabular}
				&
	\begin{tabular}{r|r|r} 
		0.183	&	0.447	&	0.458 
	\end{tabular}
				&
	\begin{tabular}{r|r|r} 
		5.067	&	3.004	&	1.845 
	\end{tabular}
	\\ \hline

6		&	

	\begin{tabular}{r|r|r} 
		0.116	&	0.430	&	0.888 
	\end{tabular}
				&
	\begin{tabular}{r|r|r} 
		0.071	&	0.207	&	0.346
	\end{tabular}
				&
	\begin{tabular}{r|r|r} 
		2.578	&	1.928	&	1.638 
	\end{tabular}
	\\ \hline

7		&	

	\begin{tabular}{r|r|r} 
		0.077	&	0.353	&	0.760
	\end{tabular}
				&
	\begin{tabular}{r|r|r} 
		0.032	&	0.130	&	0.218
	\end{tabular}
				&
	\begin{tabular}{r|r|r} 
		1.711	&	1.583	&	1.402
	\end{tabular}
	\\ \hline

8		&	

	\begin{tabular}{r|r|r} 
		0.080	&	0.281	&	0.698 
	\end{tabular}
				&
	\begin{tabular}{r|r|r} 
		0.035	&	0.058	&	0.156
	\end{tabular}
				&
	\begin{tabular}{r|r|r} 
		1.778	&	1.260	&	1.288
	\end{tabular}
	\\ \hline

9		&	

	\begin{tabular}{r|r|r} 
		0.083	&	0.262	&	0.698 
	\end{tabular}
				&
	\begin{tabular}{r|r|r} 
		0.038	&	0.039	&	0.156 
	\end{tabular}
				&
	\begin{tabular}{r|r|r} 
		1.844	&	1.175	&	1.288
	\end{tabular}
	\\ \hline
	
$\lambda_B$		&	

	\begin{tabular}{r|r|r} 
		0.045	&	0.223	&	0.542 
	\end{tabular}
				&

				&

	\\ \hline
\end{tabular}
}

\end{table}



\subsection{Decision Theoretic Model}
Experiments 1 and 2 consider an attacker that gives up after a fixed number $B$ of incorrect guesses. While this model may be appropriate in some scenarios where the attacker is eventually locked out, it does not model scenarios in which guessing is throttled using CAPTCHA puzzles (e.g., Gmail). This approach has the advantage in that legitimate users will never be locked out (at worst they will be bothered to solve a CAPTCHA puzzle). In this section, we model a rational online attacker who will select a threshold $B^{opt}$ which maximizes his expected gain (expected reward minus expected guessing costs).  In other words, the attacker will continue attacking as long as marginal reward exceeds marginal guessing costs.

\paragraph{Marginal Guessing Reward} Suppose that a rational attacker has value $v$ for a cracked password. The attacker's expected reward is $v$ times the probability he successfully cracks the password. If, as in experiment 1 (resp. experiment 2), the attacker doesn't (resp. does) know the password length $\ell$ then the expected reward after $B$ guesses is $R(v,B) \doteq v \lambda_B$ (resp. $R^\ell(v,B) \doteq v \lambda_{B,\ell}^*$). The marginal reward of one more guess when the attacker doesn't (resp. does) know the password length is $MR(v,B) = R(v,B+1)-R(v,B) = vp_{B+1}$ (resp. $MR^\ell(v,b) = R^\ell(v,B+1)-R^\ell(v,B) = v p_{B+1}^\ell$).

\paragraph{Marginal Guessing Costs} Assume that the cost of each additional password guess is $k$ (e.g., the amortized cost of paying a human to solve one more CAPTCHA puzzle). If, as in experiment 1 (resp. experiment 2), the attacker doesn't (resp. does) know the password length $\ell$ then the expected guessing cost is 
\[ C(k,B) = (1-\lambda_B) Bk +  k \sum_{i=1}^B i\times p_i \ , \] or if password length is known \[C^\ell(k,B) = (1-\lambda_{B,\ell}^*) Bk +  k \sum_{i=1}^B i\times p_i^\ell. \]
To understand this formula, we first observe that the attacker incurs maximum guessing cost $Bk$ when he fails to crack the password, which happens with probability $1-\lambda_B$ (resp.  $1-\lambda_{B,\ell}^*$) when the attacker is not told (resp. is told) the password length $\ell$. If the attacker is successful on guess $i < B$ then the attacker only incurs cost $ik$ and this happens with probability $p_i$ (resp.  $p_i^\ell $) when the attacker is not told (resp. is told) the password length $\ell$.

\paragraph{Attacker Gain} We $G(v,k,B) \doteq R(v,B)-C(k,B)$ (resp. $G^\ell(v,k,B) \doteq R^\ell(v,B)-C^\ell(k,B)$) to denote the attackers expected gain (guessing reward minus guessing cost) when  the attacker is not told (resp. is told) the password length $\ell$. If the attacker is rational the attacker will select a guessing threshold $B$ which maximizes his gain. We use $B_{v,k}^{opt} \doteq \arg\max_{B} G(v,k,B)$ resp. $B_{v,k,\ell}^{opt} \doteq \arg\max_{B} G^\ell(v,k,B)$ to denote the attackers optimal guessing threshold when not told (resp. is told) the password length $\ell$. We use $G(v,k) \doteq G(v,k,B_{v,k}^{opt} )$ denotes the  expected gain of a rational attacker in experiment 1. Finally, we use  \[G^*(v,k) \doteq \sum_{\ell} \Pr[ \mathcal{P}^\ell ]   G^\ell(v,k,B_{v,k,\ell}^{opt}) \] 
the  expected gain of a rational attacker in experiment 2.

\begin{table}

\caption{Attacker gains for various account value to marginal guessing cost ratios} \label{tab:econadvG}
\resizebox{\columnwidth}{!}{%
\begin{tabular}{|l|c|c|c|}
\hline
\multicolumn{4}{|c|}{LinkedIn} \\
\hline
Adversary type	&	Hacker	& 	Criminal	& Nation-state  \\ \hline

$v/k$ ratio	&	\begin{tabular}{l|l} $10^2$	& $10^3$ \end{tabular}
			&	\begin{tabular}{l|l} $10^4$	& $10^5$ \end{tabular}
			&	\begin{tabular}{l|l} $10^6$	& $10^7$ \Warning \end{tabular} \\ \hline
			
$ G^* / k$	
			& \begin{tabular}{r|r} 0.511	& 17.89 \end{tabular}
			& \begin{tabular}{r|r} 605.63	& 14652 
\end{tabular}
			& \begin{tabular}{r|r} 296826	& 8024681 \end{tabular} \\ \hline
			
$G / k$	
			& \begin{tabular}{r|r} 0.000	& 5.803 \end{tabular}
			& \begin{tabular}{r|r} 187.67	& 7448.19 \end{tabular}
			& \begin{tabular}{r|r} 165335	& 3085617 \end{tabular} \\ \hline
		
$\left(G^*-G\right) / k$	
			& \begin{tabular}{r|r} 0.511	& 12.097 \end{tabular}
			& \begin{tabular}{r|r} 417.96	& 7203.81 \end{tabular}
			& \begin{tabular}{r|r} 131491	& 4939064 \end{tabular} \\ \hline
			
$G^*/G$	
			& \begin{tabular}{r|r} $\infty$	& 3.083 \end{tabular}
			& \begin{tabular}{r|r} 3.227	& 1.967 \end{tabular}
			& \begin{tabular}{r|r} 1.795	& 2.600 \end{tabular} \\ \hline

\multicolumn{4}{|c|}{Rockyou} \\
\hline
Adversary type	&	Hacker	& 	Criminal	& Nation-state \Warning \\ \hline

$v/k$ ratio	&	\begin{tabular}{l|l} $10^2$	& $10^3$ \end{tabular}
			&	\begin{tabular}{l|l} $10^4$	& $10^5$ \end{tabular}
			&	\begin{tabular}{l|l} $10^6$	& $10^7$ \end{tabular} \\ \hline
			
$G^* / k$	
			& \begin{tabular}{r|r} 0.959	& 30.168 \end{tabular}
			& \begin{tabular}{r|r} 1225.5	& 27053 \end{tabular}
			& \begin{tabular}{r|r} 552447	& 9.5E6 \end{tabular} \\ \hline
			
$G / k$	
			& \begin{tabular}{r|r} 0.000	& 12.141 \end{tabular}
			& \begin{tabular}{r|r} 428.182	& 14576 \end{tabular}
			& \begin{tabular}{r|r} 297611	& 6.7E6 \end{tabular} \\ \hline
		
$\left(G^*-G\right) / k$	
			& \begin{tabular}{r|r} 0.959	& 18.026 \end{tabular}
			& \begin{tabular}{r|r} 797.31	& 23577 \end{tabular}
			& \begin{tabular}{r|r} 254836	& 2.8E6 \end{tabular} \\ \hline
			
$G^*/G$	
			& \begin{tabular}{r|r} $\infty$	& 2.485 \end{tabular}
			& \begin{tabular}{r|r} 2.862	& 1.856 \end{tabular}
			& \begin{tabular}{r|r} 1.856	& 1.422 \end{tabular} \\ \hline

\end{tabular}	
}


\end{table}

{\bf Attacker's Monetary Benefit when Learning Password Lengths:} We remark that $G^*(v,k)-G(v,k)$ denotes the expected (per user) benefit to the online attacker when learning the password length $\ell$. If this benefit $\#(\mbox{AttackedUsers} \times \left( G^*(v,k)-G(v,k)\right)$ exceeds the cost of eavesdropping on network traffic then it will be worthwhile for the attacker to exploit the password length-leakage attacks described earlier. Table \ref{tab:econadvG} plots the value $G^*(v,k)-G(v,k)$ for different $v/k$ ratios. Tables showing the optimal thresholds $B^{OPT}_{...}$ which maximize gains at various $v/k$ ratios can be found in the Appendix in Table 8. From Table \ref{tab:econadvG} we can see that the monetary benefit of learning password length can be quite substantial. The increased monetary benefit might entice additional criminals to entire the password cracking game provided that increased monetary benefit outweighs the cost of eavesdropping and linking IP addresses to user ids. As Rob Joyce, Chief of Tailored Access Operations at NSA, said, ``Don't assume a crack is too small to be noticed, or too small to be exploited.''~\cite{Zetter}

\paragraph{Example:} Suppose that it costs $k=\$0.001$ (the approximate cost of paying a human to solve a CAPTCHA puzzle) per password guess and the value of each cracked password to a criminal is $v = 10^5 k = \$100$. We have a $v/k$ ratio of $10^5$, and can look up the value $\left(G^* - G / k \right) \approx 7203$ in Table \ref{tab:econadvG} (using estimates derived from the  LinkedIn frequency corpus). We can convert this to a monetary value by multiplying by our $k$ value. Here we see that knowing password length gains us an average of $\$7.23$ per password. Say an attacker is targeting $100$ people. So long as the cost of sniffing network traffic is $< \$723$ and it is possible for the attacker to link user ids with each IP address it is worth the attacker's time to run the attack.

\begin{table}

\caption{Advantages for several guessing limits}\label{tab:econadvLambd}
\resizebox{\columnwidth}{!}{%
\begin{tabular}{|l|c|c|c|}
\hline
\multicolumn{4}{|c|}{LinkedIn} \\
\hline
Adversary type	&	Hacker	& 	Criminal	& Nation-state \\ \hline

Guess limit	&	\begin{tabular}{l|l} $10^2$	& $10^3$ \end{tabular}
			&	\begin{tabular}{l|l} $10^4$	& $10^5$ \end{tabular}
			&	\begin{tabular}{l|l} $10^6$	& $10^7$ \Warning \end{tabular} \\ \hline
$\overline{\lambda}^*_{v,k}$	
			& \begin{tabular}{r|r} 0.007	& 0.028 \end{tabular}
			& \begin{tabular}{r|r} 0.090	& 0.191 \end{tabular}
			& \begin{tabular}{r|r} 0.378	& 1.000 \end{tabular} \\ \hline			
$\overline{\lambda}^*_{v,k}-\overline{\lambda}_{v,k}$	
			& \begin{tabular}{r|r} 0.000	& 0.019 \end{tabular}
			& \begin{tabular}{r|r} 0.055	& 0.084 \end{tabular}
			& \begin{tabular}{r|r} 0.164	& 0.619 \end{tabular} \\ \hline
			
$\overline{\lambda}^*_{v,k}/\overline{\lambda}_{v,k}$	
			& \begin{tabular}{r|r} $\infty$	& 3.182 \end{tabular}
			& \begin{tabular}{r|r} 2.586	& 1.785 \end{tabular}
			& \begin{tabular}{r|r} 1.766	& 2.625 \end{tabular} \\ \hline

\multicolumn{4}{|c|}{Rockyou} \\
\hline
Adversary type	&	Hacker	& 	Criminal	& Nation-state \Warning \\ \hline

Guess limit	&	\begin{tabular}{l|l} $10^2$	& $10^3$ \end{tabular}
			&	\begin{tabular}{l|l} $10^4$	& $10^5$ \end{tabular}
			&	\begin{tabular}{l|l} $10^6$	& $10^7$ \end{tabular} \\ \hline
$\overline{\lambda}^*_{v,k}$	
			& \begin{tabular}{r|r} 0.014	& 0.052 \end{tabular}
			& \begin{tabular}{r|r} 0.177	& 0.363 \end{tabular}
			& \begin{tabular}{r|r} 0.823	& 1.000 \end{tabular} \\ \hline			
$\overline{\lambda}^*_{v,k}-\overline{\lambda}_{v,k}$	
			& \begin{tabular}{r|r} 0.014	& 0.034 \end{tabular}
			& \begin{tabular}{r|r} 0.104	& 0.163 \end{tabular}
			& \begin{tabular}{r|r} 0.446	& 0.000 \end{tabular} \\ \hline
			
$\overline{\lambda}^*_{v,k}/\overline{\lambda}_{v,k}$	
			& \begin{tabular}{r|r} $\infty$	& 2.874 \end{tabular}
			& \begin{tabular}{r|r} 2.417	& 1.817 \end{tabular}
			& \begin{tabular}{r|r} 2.183	& 1.000 \end{tabular} \\ \hline

\end{tabular}	
}


\end{table}

\begin{table}
\caption{Cracking estimates for attacks over time (LinkedIn)} \label{timeattack}
\resizebox{\columnwidth}{!}{%
\begin{tabular}{|c|c|c|c|}
\hline
\multicolumn{4}{|c|}{LinkedIn} \\
\hline
Days & 1 guess/day & 10 guesses/day & 100 guesses/day (LinkedIn) \\ \hline
30		&	0.04(0.02)	&	0.08(0.05)		&	0.16(0.10)	\\ \hline
90		&	0.06(0.03)	&	0.11(0.07)		&	0.21(0.14)	\\ \hline
180		&	0.7(0.04)	&	0.14(0.09)		&	0.24(0.16)	\\ \hline
360		&	0.9(0.5)	&	0.17(0.11)		&	0.28(0.19)	\\ \hline

\multicolumn{4}{|c|}{Rockyou} \\
\hline

Days & 1 guess/day & 10 guesses/day & 100 guesses/day (Rockyou) \\ \hline
30		&	0.06(0.03)	&	0.13(0.07)		&	0.26(0.16)	\\ \hline
90		&	0.09(0.04)	&	0.19(0.11)		&	0.32(0.22)	\\ \hline
180		&	0.11(0.06)	&	0.23(0.14)		&	0.37(0.26)	\\ \hline
360		&	0.14(0.08)	&	0.27(0.17)		&	0.43(0.30)	\\ \hline

\end{tabular}
}

\end{table}

{\bf Number of Compromised User Accounts:} We now seek to quantify the damages of leaking password lengths to a rational attacker. In particular, we use $\overline{\lambda}_{v,k} = \lambda_{B_{v,k}^{opt}}$ (resp. $\overline{\lambda}_{v,k,\ell}^* = \lambda_{B_{v,k,\ell}^{opt},\ell}^*$) to denote the probability a rational value $v$ attacker succeeds without (resp. with) knowledge of the password length $\ell$ and given guessing costs $k$. We use  $\overline{\lambda}_{v,k}^* = \sum_{\ell} \Pr[ \mathcal{P}^\ell ]   \overline{\lambda}_{v,k,\ell}^*$ to denote the probability that a password is cracked in experiment 2 (We will also use $\overline{\lambda}_{v,k,\ell} = \lambda_{B_{v,k}^{opt},\ell}$ to denote the probability that an attacker cracks the user's password without knowledge of password length conditioning on the event that the user's password has length $\ell$.). Finally, we note that $\overline{\lambda}_{v,k}^* - \overline{\lambda}_{v,k}$ denotes the increase in the attacker's success rate when the attacker learns the password length. 

Appendix Table 7 compares the attackers success rate with $\left(\overline{\lambda}_{v,k}^*\right)$ and without $\left(\overline{\lambda}_{v,k,\ell}\right)$ knowledge of the password length for various $v/k$ ratios, and Table 4 compares the attacker's success rate conditioning on the event that the  the password has length $\ell$ (Appendix Table 8 shows the optimal thresholds $B^{OPT}_{v,k}$ as well as $B^{OPT}_{v,k,\ell}$ for various lengths $\ell$). These tables show that a rational attacker will crack many more passwords when given the password length. For example, a criminal attacker with $v/k = 10^5$ who knows the password length will crack over $19\%$ of targets from the Linkedin data compared to just $10.7\%$ of targets without knowledge of password length. As another example consider a hacker using the Rockyou data with $v/k=10^2$. If the attacker does not know the password length then his optimal strategy is to give up immediately without attempting any password guesses. However, if the attacker does know the length then he will crack about $1.4\%$ of passwords. We once again wish to stress the $\Warning$ symbol towards higher $v/k$ ratios, which denotes situations where overestimates are likely - especially where values show the adversary would guess $100\%$ of passwords.   

{\bf Online time-delayed attacks} The models we have introduced are based on the notion that an attacker can continuously try passwords. However, in many situations, there is some sort of lockout that limits the number of attempts that can be made. In this case, the adversary can run an attack over time to bypass lockout mechanisms. Rather than being rate limited by a service like CAPTCHA, a set number of guesses may be run per day. The models introduced also provide insight into these types of attacks. To provide an idea of what sort of advantage an adversary may have in this case, we take Brostoff and Sasse's recommendations that 10 attempts should be allowed \cite{brostoff2003ten}, however, we note that Bonneau and Preibusch found that the vast majority of sites they surveyed allowed over 100 guesses with no restrictions \cite{bonneau2010password}. In addition, the National Institute of Standards and Technology recommendations allow for no more than 100 consecutive failed login attempts before a lockout \cite{nistpassword}. Table 5 shows the estimated proportion of passwords that would be cracked over set periods of time given 1, 10, and 100 daily guesses.

{\bf Notation Remark: } We use the notation $\overline{\lambda}_{...}$ when considering a rational (utility optimizing) attacker and $\lambda_{\ldots}$ when considering an attacker that is either locked out or gives up after a fixed number of guesses. We also remark that we use $*$ to indicate an experiment in which the attacker knows the length, and we use $\ell$ when conditioning on the event that the user's password has length $\ell$ (whether or not this length is known to the attacker). Thus, $\overline{\lambda}_{v,k,\ell}$ denotes the conditional success rate of a rational attacker who does not knows the password length conditioning on the event that the users password is length $\ell$ and $\lambda_{B}^*$ denotes the success rate of an attacker who knows the password length and tries exactly $B$ guesses (without conditioning on the event that the users password is a particular length). 
\paragraph{Differences Between Our Model and~Previous Work\cite{blockiCASH2016,SP:BloHarZho18}} Blocki and Datta~\cite{blockiCASH2016} and Blocki et al.~\cite{SP:BloHarZho18} to model an offline password attacker. By contrast, we consider online attacks.  Blocki and Datta~\cite{blockiCASH2016} were focused on predicting $\overline{\lambda}_{v,k}$ to estimate the probability that a rational attacker cracks each password. While we are still interested in this quantity we are also interested in understanding the attackers gain so that we can predict whether or not it is worthwhile for the attacker to eavesdrop on encrypted network traffic. 


\section{Solutions}
At a conceptual level, the solution to the problem of password length leakage is straightforward. Make sure that passwords are always padded before they are encrypted. However, at a practical level, the problem defies an easy solution because any solution will require updates to web pages, browsers, and/or servers. We contend that a large part of the problem is that the responsibility of identifying and padding length-sensitive data such as passwords has been largely pushed to the application developer. 

 The issue is increasingly pressing as the Internet increasingly moves to AES-GCM with the advent of QUIC and TLS 1.3. In 2015 CBC mode was previously one of the most popular modes of operation in TLS connections~\cite{TLSDeployment}, but as of 2018, over 80\% of connections use AES-GCM. While AES-GCM has many advantages over CBC mode in terms of both efficiency and security\footnote{As an Authenticated Encryption with Associated Data cipher AES-GCM provides {\em both} confidentiality and integrity. By contrast, CBC mode only provides confidentiality meaning that additional steps (e.g., MACs, signatures) must be taken to guarantee message integrity --- a step that turns to be quite challenging to get right in practice. There have been several attacks against TLS implementations of CBC mode including BEAST (CVE-2011-3389), LUCKY13 \cite{SP:AlFPat13} and  POODLE \cite{poodle}.}, CBC mode provides natural padding i.e., messages are padded to 16-bytes blocks in CBC mode while where there is a 1-1 correspondence between plaintext and ciphertext length in AES-GCM. 

Our immediate goal is to increase security for individual privacy without requiring new transport protocol design or changing cipher suites. Our solutions are focused around the Hypertext Transfer Protocol layer (HTTP) to limit changes that could disrupt this eco-system: 1) For HTTP 1.1 developers, we outline a framework to change web pages alleviating some changes to back end servers.  2) For future development, we recommend developers use the HTTP/2 standard and use its extension to implement padding. 3) We also advocate that W3C be updated to specify that password fields must be padded by default.

Our immediate goal is to increase security for individual privacy without requiring new protocol design or changing cipher suites and to do this, we believe developing a software solution at the Hypertext Transfer Protocol layer is the best way to achieve these goals. 

It is important to develop a solution with compatible technologies.  Since TLS version 1.2 and QUIC have been in existence since at least 2013, we believe that using a client and server software solution in JavaScript with the AJAX framework and JSON files will meet our compatibility requirements for most servers and web browsers.   This will allow for an immediate fix to increase individual privacy and security to websites containing sensitive identifiers commonly found in web forms.   Our general scheme is as follows with three password transformation processes (hashing, padding) options:
	The server will first identify web forms with sensitive form fields and send the web form with a client-side script.  The client-side script will transform the password by either hashing it or padding it. The client will submit the web form with the transformed password to the server.  The server will then decrypt the information from the previously established secure channel (i.e. TLS version 1.2, 1.3 or QUIC).  Finally, the server will transform the padded password back into the real password (or work directly with the hashed password).  

{\bf Hashing.}  Hashing the password on the client-side is the most straight-forward approach. Instead of sending the server the user's actual password $pwd_u$ we would instead send the hash value $H(pwd_u)$ or $H(u,pwd_u)$. Hashing the password has the benefit of being easy to implement on client side and immediately protects the password on the client and server to ensure some security protections from both online and offline attacks. Intel and AMD processors both support hardware acceleration for the SHA family of hashing algorithms via the Intel SHA Extensions to the x86 instruction set architecture so the solution could be made to be efficient. However, this approach would require server-side changes\footnote{Instead of storing the salted hash of a user's password  the server would have to store the hash of the hash e.g., $\left(u,s_u,H(s_u, SHA3(u, pwd_u))\right)$ instead $(u,s_u,H(s_u, pwd_u))$. Until the user $u$ provides the password $pwd_u$ there is no way to directly obtain the new record  $\left(u,s_u,H(s_u, SHA3(u, pwd_u))\right)$ from the current record $(u,s_u,H(s_u, pwd_u))$ which is already stored on the server.}. 

{\bf Note:} It would be recommended that the server select a strong password hash function $H$ which is moderately expensive to compute to protect against offline attackers~\cite{nistpassword}. PBKDF2 uses hash iteration to increase guessing costs but is not memory-hard. Blocki et al.~\cite{SP:BloHarZho18} argued that functions like PBKDF2 which are not memory hard provide insufficient protection for low-entropy secrets such as passwords. While the National Institute of Standards and Technology has yet not vetted and approved any memory hard function~\cite{nistpassword}, the Memory Hard Function Argon2~\cite{biryukov2016argon2} was selected as the winner of the password hashing competition in 2015~\cite{EPRINT:Wetzels16}. 


 The use of a memory-hard function such as Argon2 would be best against offline attacks while protecting the length.  A memory hard function has yet to be vetted and approved by the National Institute of Standards and Technology.  Just In Time hashing (JIT hashing)  is an another alternative to hashing that does not deter from the online experience of the user. As soon as the user types in the first character(s) of their password, the JIT algorithm fills in memory with the hash values for the characters the user has typed \cite{harsha2018just}.  JIT is not widely implemented in web scripting languages and may be more difficult for common developers to implement widely.

{\bf Padding.}   Similar to hashing, padding can help hide the length of the password. TLS (v1.2 and v1.3) and QUIC both support optional padding parameters. In our solutions, the client-side script provided by the server will use this option to ensure that the password field is padded. This approach has the benefit of requiring minimal server-side changes (apart from identifying web forms with passwords and sending the appropriate client-side script) as TLS/QUIC will automatically discard the padding on the server. The process is opaque to both the developer and the user.

{\bf Note: } Block cipher modes such as CBC would provide natural padding, but CBC mode does not guarantee message integrity and there have been several attacks against TLS implementations of CBC mode including BEAST (CVE-2011-3389), LUCKY13 \cite{SP:AlFPat13} and  POODLE \cite{poodle}. We also find that GCM is much more efficient due to native acceleration.  

{\bf HTTP/2: }For developers moving to HTTP/2, we recommend using the HTTP/2 padding flag contained with the DATA frames.  DATA frames convey arbitrary, variable-length sequences of octets associated with a stream.  One or more DATA frames are used to carry HTTP request or response payloads.  DATA frames may also contain padding. Padding can be added to DATA frames to obscure the size of messages.  It should be noted padding within HTTP/2 is not intended as a replacement for general purpose padding such as with TLS version 1.2.   It is recommended intermediaries should retain padding for data frames and drop padding for HEADERS and PUSH PROMISE frames because an intermediary can change the protections padding provides. This offers minimal protection for HTTP/2 environments. 


{\bf Long-Term Solutions.} As for less immediate solutions, we suggest the consideration of changes to future versions of standards and protocols. In the case of W3 standards, they specify a special input type for passwords in web forms. For example, the login form for a well known banking website includes the following code snippet:
\begin{verbatim}
<input id="usr_password_home" ... 
type="password" maxlength="32"...> 
\end{verbatim}
As the above example illustrates a password input box allows for an (optional) property called maxlength. We advocate for an update to W3 standards to include a boolean ``length-sensitive'' attribute for all input types. If this attribute is set to true then the browser will pad the input value to maxlength, or in the event that the maxlength attribute is not specified the browser could pad the input value, e.g., to ensure that the plain text length is a multiple of $30$. An update to W3 standards would be easy to implement at the browser level. Ideally, it would be nice to set the ``length-sensitive'' attribute to true by default for the password input type. The advantage of setting the attribute to true by default is that we would no longer rely on a web developer to remember to set this attribute to true for passwords. On the downside setting the attribute to true by default may create backward compatibility issues with any web server using an older version of the W3 standard. 

To defend against password attacks (e.g., password spraying), we suggest enabling multi-factor authentication for all internet facing applications.  Organizations should also align password policies with NIST guidelines~\cite{nistpassword}.  Some companies are adding new detection and prevention tools such as IP lockout when analyzing sign-ons to the same account from different IP addresses \cite{simons_2018}.

\section{Conclusions} \label{sec:conclusions}
In TLS 1.3 and TLS 1.2 (some ciphers) there is a 1-1 relationship between the length of a ciphertext and the length of the corresponding plaintext. The responsibility of identifying and padding length sensitive data is pushed to the application developer. We conducted an observational study of AES-GCM traffic (the most commonly used cipher in TLS) which uncovered a widespread failure to pad passwords. In particular, we found multiple high profile instances where password lengths can be directly inferred from encrypted web traffic. If an eavesdropping attacker is able to link the source IP address with a particular user name (e.g., via unencrypted traffic) then the attacker can directly infer the length of that user's password. We used a decision-theoretic model to analyze the advantage a password attacker obtains by learning the length of a user's password. Our analysis shows that the advantage is substantial.

While there are good metrics about how much a particular cipher reduces latency, there are fewer reliable models to help security professionals quantify how many users might be susceptible to intrusions and from which class of intruders (i.e. nation-states, criminals, or hackers).   Without this information, it will be difficult for security professionals to make informed decisions about the trade-offs between speed and security.  As a necessary first step, this research begins to quantify the cost an intruder would incur to perform an online attack.  This helps better understand which intruders might be willing to endure the cost to intrude on when conducting a campaign against a set of targeted accounts.  As the push for faster security transport security protocols continues to grow, researchers will be challenged to balance the need for speed and security. Ultimately, this research adds value by helping security researchers better quantify and compare the specific trade-offs between speed and security for a particular cipher suite used for a security transport protocol.

\subsection{Future Work}
We have begun to quantify the risks of leaking password lengths. There is a need to quantify the risks of leaking plain text lengths in other contexts (e.g., short chat communications). In general, it would be helpful to formulate clear guidelines to help developers evaluate when plain text length should be viewed as a sensitive attribute. When plain text length is sensitive there is a need to provide developers with an easy method to obfuscate sensitive data lengths, and there is a need to develop automated tools which could audit code and identify instances where length-sensitive data might not be hidden.

\subsection{Recommendations for Users} We offer the following suggestions to users to protect themselves against password length leakage attacks\footnote{Even in the absence password length leakage attacks the following advice can help a user to secure his accounts. However, the advice takes on a greater urgency due to the stronger threat of online attackers.}. First, enable two-factor authentication whenever possible. Strong two-factor authentication will prevent an attacker from mounting an online attack whether or not the attacker knows the length $\ell$ of your password. Second, we recommend that users select strong passwords which don't occur in a password cracking dictionary. We recognize that this advice, while easy to give, can be challenging to follow since users typically have many password protected accounts~\cite{florencio2014password}. However, the additional risks from password length leakage may justify the extra effort for many users. Mnemonic techniques~\cite{AC:BloBluDat13,CCS:YLCXP16} and spaced repetition~\cite{bonneau2014towards,NDSS:BKCD15} may help to reduce the extra user burden. Finally, users could also begin to use a password manager to generate unique passwords from one master password. This solution potentially reduces user burden since the user will now only need to remember a few master passwords. Second, the length of the derived password for each domain is not necessarily correlated with the length of the user's master password.  For example, PwdHash~\cite{ross2005stronger} derives a unique fixed-length password for each domain based on the cryptographic hash of the user's master password along with the domain itself.

\section{Acknowledgments}

The authors wish to express their appreciation for the support that they received from the INSuRE (Information Security Research and Education) program as INSuRE provided the venue for the initial inquiry; INSuRE has been supported in part by the National Science Foundation under a variety of grants including an Eager grant (1344369) as well as supplements to several SFS and other grants (1241576, 1433753, 1433690, 1241668, 1433795, 1241576, 1241668). INSuRE also received support from the National Security Agency via grants H98230-15-1-0298, H98230-15-1-0299, H98230-15-1-0300, and H98230-17-1-0314. The work was also supported in part by NSF CNS\#1704587 and Ben Harsha was supported by a Rolls Royce Doctoral Fellowship. We would also like to extend a special thanks to KoreLogic for their assistance with the LinkedIn password frequency corpus.

\bibliographystyle{ACM-Reference-Format}
\bibliography{abbrev3,crypto,biblio}

\appendix

\section{Notation Guide}\label{sec:notation}

Here we include a brief summary of the notation used in our analysis. This is meant as a quick reference for intuition - for full definitions see the main body of the paper.

\begin{itemize}
	\item[$p_i$] The proportion of users who selected the $i$'th most common password
	\item[$p_i^\ell$] The proportion of users who selected the $i$'th most common password of length $\ell$
	\item[$B$] The number of guesses an adversary is making in a password-guessing attack
	\item[$\lambda_B$] The proportion of passwords guessed in some attack with B guesses. E.g. $\lambda_{10}$ is the proportion of passwords an adversary would guess with no additional information.
	\item[star (*)] Indicates attacker knowledge of password length. No star means the attacker does not know password length, and a star means they do. E.g. $\lambda_B^*$ is the proportion of passwords guessed in an attack where the attacker knows password length.
	\item[$\ell$] Means the variable is only considering the list of passwords of length $\ell$. E.g. $\lambda_{B,\ell}$ is the proportion of passwords of length $\ell$ guessed in $B$ guesses.
	\item[$v$] The value of a guessed password
	\item[$k$] The cost of guessing a password
	\item[$R(v,B)$] The expected reward an attacker would get in an attack with $B$ guesses against a password with value $v$.
	\item[$C(k,B)$] The expected cost an attacker must pay to make $B$ guesses on a password list, with each guess costing $k$.
	\item[$G(v,k,B)$] The expected gain (i.e. Reward - Cost) of an attacker making $B$ guesses with password value $v$ and cost of each guess $k$.
	\item[$B^{OPT}_{v,k}$] The value of $B$ that maximizes $G$ for some list of passwords, value $v$, and guess cost $k$.
	\item[Bar (e.g. $\overline{\lambda}$)] Number of passwords guessed when using the relevant $B^{OPT}$ value.
\end{itemize}

\section{Alexa TOP 100}
For the following Top Alexa 100, these websites used AES-GCM as one of the potential cipher suites.  In all cases, these websites did not offer sufficient additional protection (i.e., padding) to obfuscate the password length. Thus, if a Top Alexa 100 website uses AES-GCM, denoted with a 1 in the table, then it was vulnerable to an attacker inferring the length for a plausible attack outlined in section \ref{sec:exploitlengthleakage}.

\begin{figure}
 \includegraphics[width=90mm,scale=0.5]{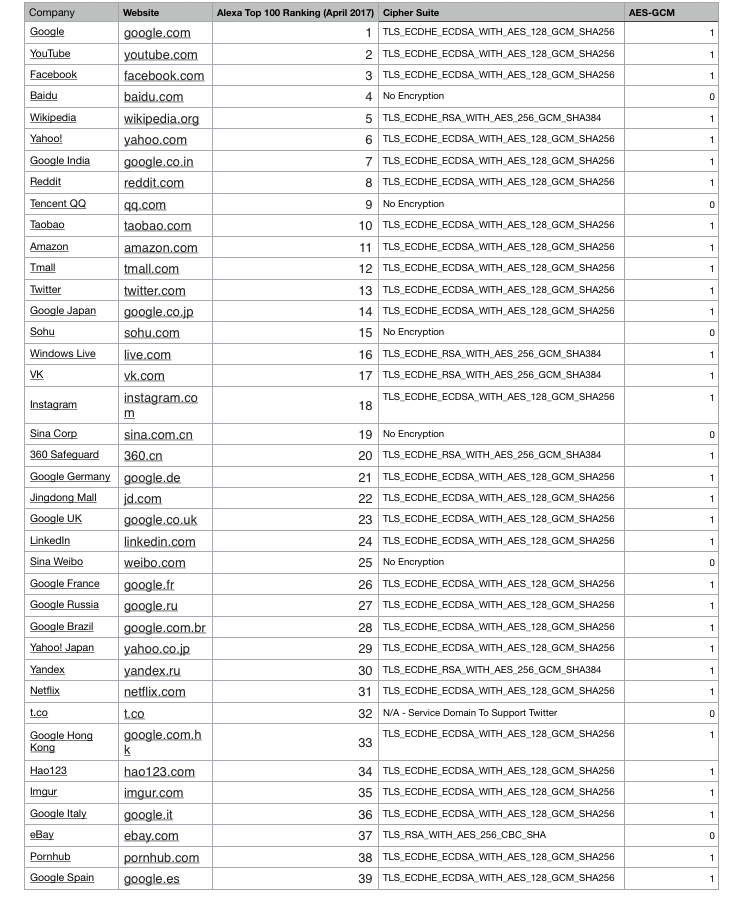}
 \caption{Top Alexa 100 using AES-GCM}
 \label{CBCvsGCM}
\end{figure}

\begin{figure}
 \includegraphics[width=90mm,scale=0.5]{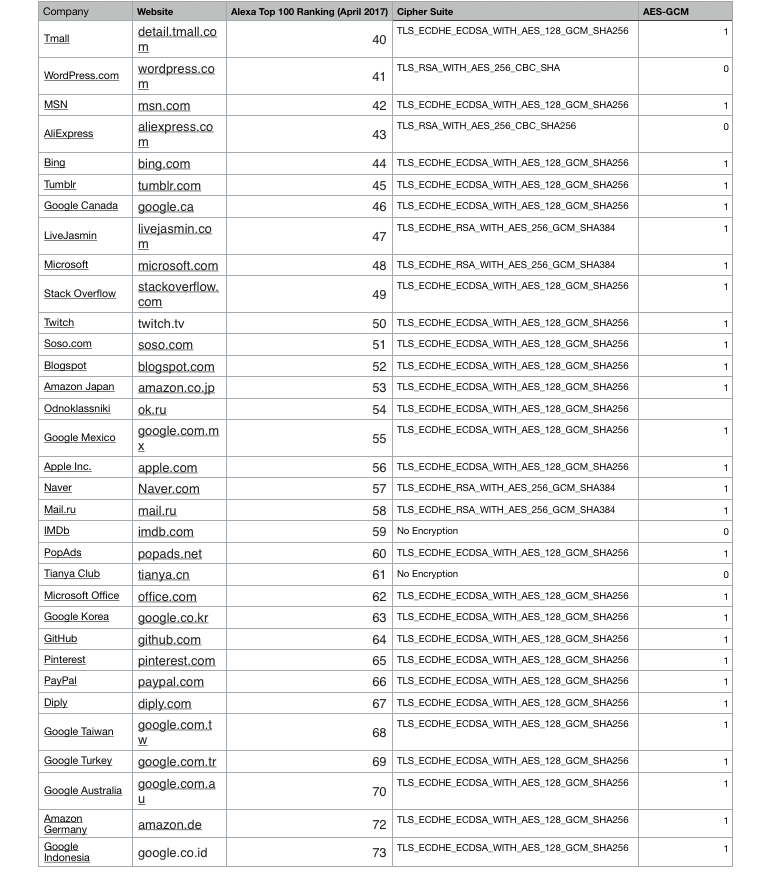}
 \caption{Top Alexa 100 using AES-GCM (Continued)}
 \label{CBCvsGCM}
\end{figure}

\begin{figure}
 \includegraphics[width=90mm,scale=0.5]{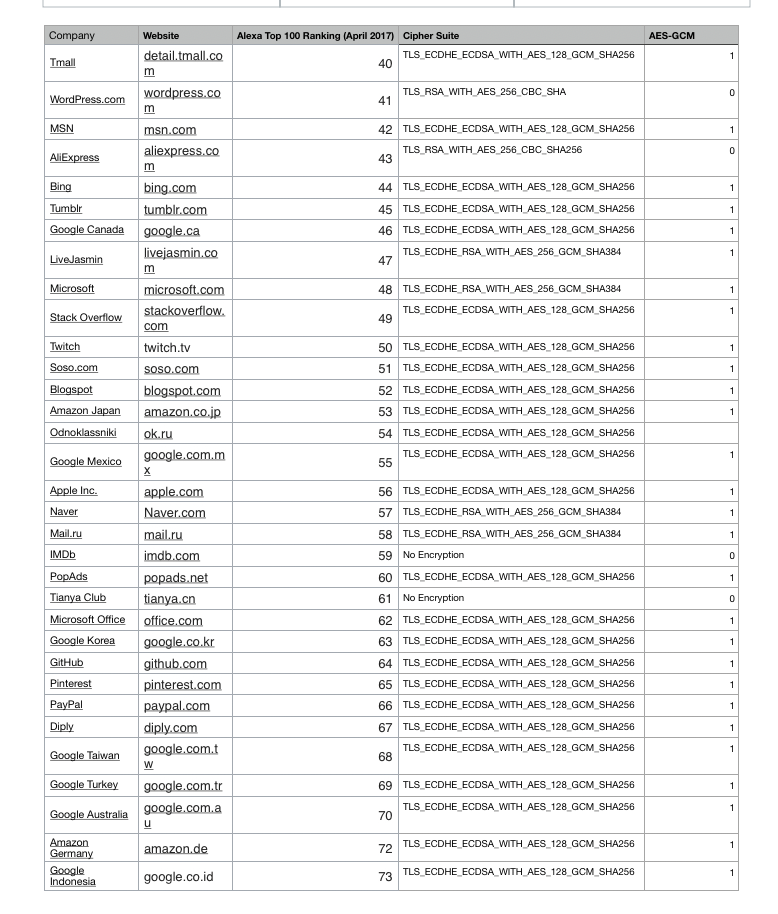}
 \caption{Top Alexa 100 using AES-GCM (Continued)}
 \label{CBCvsGCM}
\end{figure}

\begin{figure}
 \includegraphics[width=90mm,scale=0.5]{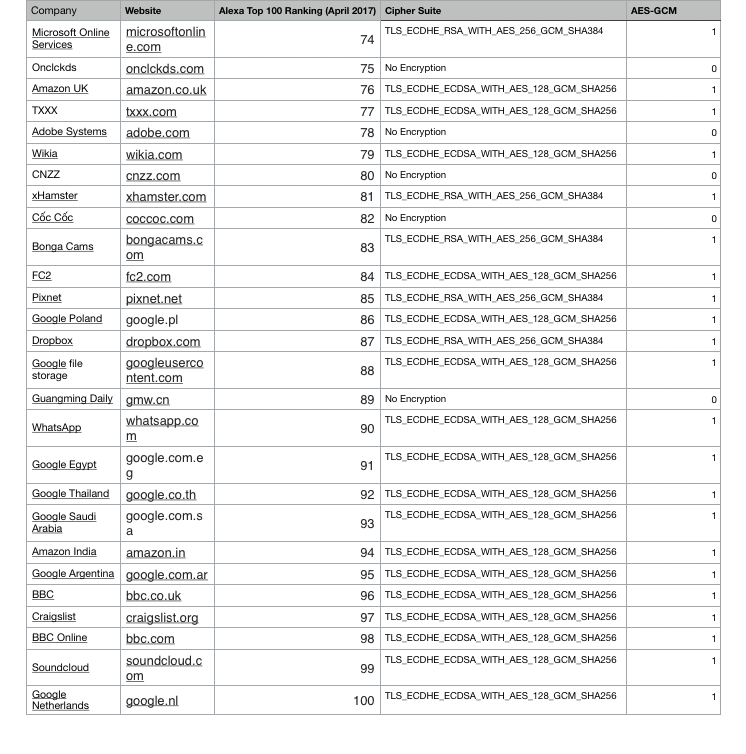}
 \caption{Top Alexa 100 using AES-GCM (Continued)}
 \label{CBCvsGCM}
\end{figure}

\subsection*{Additional tables}
\begin{table}[h]
\caption{Attacker Conditional Success Rate at Various Guessing Limits with and without knowledge of the password length}\label{tab:limitstats}

\centering
\resizebox{\columnwidth}{!}{%
\begin{tabular}{|l|c|c|c|c|}
\hline
Lengths		&	$\lambda_{B,\ell}^*$	&	$\lambda^*_{B,\ell} - \lambda_{B,\ell}$ 	&	$\lambda^*_{B,\ell}  / \lambda_{B,\ell}$ \\ \hline

limit		&	

	\begin{tabular}{r|r|r}
		$10^2$	&	$10^4$	&	$10^6$\Warning
	\end{tabular}
				&
	\begin{tabular}{c|c|c} 
		$10^2$	&	$10^4$	&	$10^6$\Warning
	\end{tabular}
				&
	\begin{tabular}{c|c|c} 
		$10^2$	&	$10^4$	&	$10^6$\Warning
	\end{tabular}
	\\ \hline

5		&	

	\begin{tabular}{r|r|r} 
		0.228	&	0.670	&	1.000 
	\end{tabular}
				&
	\begin{tabular}{r|r|r} 
		0.183	&	0.447	&	0.458 
	\end{tabular}
				&
	\begin{tabular}{r|r|r} 
		5.067	&	3.004	&	1.845 
	\end{tabular}
	\\ \hline

6		&	

	\begin{tabular}{r|r|r} 
		0.116	&	0.430	&	0.888 
	\end{tabular}
				&
	\begin{tabular}{r|r|r} 
		0.071	&	0.207	&	0.346
	\end{tabular}
				&
	\begin{tabular}{r|r|r} 
		2.578	&	1.928	&	1.638 
	\end{tabular}
	\\ \hline

7		&	

	\begin{tabular}{r|r|r} 
		0.077	&	0.353	&	0.760
	\end{tabular}
				&
	\begin{tabular}{r|r|r} 
		0.032	&	0.130	&	0.218
	\end{tabular}
				&
	\begin{tabular}{r|r|r} 
		1.711	&	1.583	&	1.402
	\end{tabular}
	\\ \hline

8		&	

	\begin{tabular}{r|r|r} 
		0.080	&	0.281	&	0.698 
	\end{tabular}
				&
	\begin{tabular}{r|r|r} 
		0.035	&	0.058	&	0.156
	\end{tabular}
				&
	\begin{tabular}{r|r|r} 
		1.778	&	1.260	&	1.288
	\end{tabular}
	\\ \hline

9		&	

	\begin{tabular}{r|r|r} 
		0.083	&	0.262	&	0.698 
	\end{tabular}
				&
	\begin{tabular}{r|r|r} 
		0.038	&	0.039	&	0.156 
	\end{tabular}
				&
	\begin{tabular}{r|r|r} 
		1.844	&	1.175	&	1.288
	\end{tabular}
	\\ \hline
	
$\lambda_B$		&	

	\begin{tabular}{r|r|r} 
		0.045	&	0.223	&	0.542 
	\end{tabular}
				&

				&

	\\ \hline
\end{tabular}
}

\end{table}

\begin{table}[h]

\caption{Cracking data for limited guessing} \label{tab:econadvLambdaTwo}
\resizebox{\columnwidth}{!}{%
\begin{tabular}{|l|c|c|c|c|}
\hline
Lengths		&	$\overline{\lambda}_{v,k,\ell}^*$	&	$\overline{\lambda}^*_{v,k,\ell} - \overline{\lambda}_{v,k,\ell}$ 	&	$\overline{\lambda}^*_{v,k,\ell}  / \overline{\lambda}_{v,k,\ell}$ \\ \hline

limit		&	

	\begin{tabular}{r|r|r}
		$10^2$	&	$10^4$	&	$10^6$ 
	\end{tabular}
				&
	\begin{tabular}{c|c|c} 
		$10^2$	&	$10^4$	&	$10^6$ 
	\end{tabular}
				&
	\begin{tabular}{c|c|c} 
		$10^2$	&	$10^4$	&	$10^6$ 
	\end{tabular}
	\\ \hline

5		&	

	\begin{tabular}{r|r|r} 
		0.060	&	0.522	&	1.000 
	\end{tabular}
				&
	\begin{tabular}{r|r|r} 
		0.060	&	0.249	&	0.315 
	\end{tabular}
				&
	\begin{tabular}{r|r|r} 
		$\infty$	&	1.912	&	1.460 
	\end{tabular}
	\\ \hline

6		&	

	\begin{tabular}{r|r|r} 
		0.034	&	0.258	&	1.000 
	\end{tabular}
				&
	\begin{tabular}{r|r|r} 
		0.000	&	0.059	&	0.190
	\end{tabular}
				&
	\begin{tabular}{r|r|r} 
		1.000	&	1.300	&	1.235 
	\end{tabular}
	\\ \hline

7		&	

	\begin{tabular}{r|r|r} 
		0.000	&	0.182	&	0.596
	\end{tabular}
				&
	\begin{tabular}{r|r|r} 
		0.000	&	0.081	&	0.132
	\end{tabular}
				&
	\begin{tabular}{r|r|r} 
		N/A	&	1.802	&	1.284
	\end{tabular}
	\\ \hline

8		&	

	\begin{tabular}{r|r|r} 
		0.000	&	0.128	&	0.500 
	\end{tabular}
				&
	\begin{tabular}{r|r|r} 
		0.000	&	0.059	&	0.132
	\end{tabular}
				&
	\begin{tabular}{r|r|r} 
		N/A	&	1.855	&	1.359
	\end{tabular}
	\\ \hline

9		&	

	\begin{tabular}{r|r|r} 
		0.019	&	0.121	&	1.000 
	\end{tabular}
				&
	\begin{tabular}{r|r|r} 
		0.019	&	0.064	&	0.863 
	\end{tabular}
				&
	\begin{tabular}{r|r|r} 
	$\infty$	&	2.122	&	7.300
	\end{tabular}
	\\ \hline
	
$\overline{\lambda}^*_{v,k}$	&	

	\begin{tabular}{r|r|r} 
		0.014	&	0.177	&	0.822 
	\end{tabular}
				&

				&

	\\ \hline
\end{tabular}
}

\end{table}

\begin{table}[h]

\caption{Optimal Attacker Guessing Limit for various account value to marginal guessing cost ratios} 
\label{tab:bopt}
\resizebox{\columnwidth}{!}{%

\begin{tabular}{|l|c|c|c|}
\hline
\multicolumn{4}{|c|}{LinkedIn} \\
\hline
Adversary type	&	Hacker	& 	Criminal	& Nation-state \\ \hline

$v/k$ ratio	&	\begin{tabular}{l|l} $10^2$	& $10^3$ \end{tabular}
			&	\begin{tabular}{l|l} $10^4$	& $10^5$ \end{tabular}
			&	\begin{tabular}{l|l} $10^6$	& $10^7$ \Warning \end{tabular} \\ \hline

$B^{OPT}_{v,k,6}$
			&	\begin{tabular}{l|l} 1	& 10 \end{tabular}
			&	\begin{tabular}{l|l} 736	& 10327 \end{tabular}
			&	\begin{tabular}{l|l} 271903	& 5418647 \end{tabular} \\ \hline
$B^{OPT}_{v,k,7}$
			&	\begin{tabular}{l|l} 0	& 4 \end{tabular}
			&	\begin{tabular}{l|l} 570	& 8818 \end{tabular}
			&	\begin{tabular}{l|l} 134534	& 6672366 \end{tabular} \\ \hline
$B^{OPT}_{v,k,8}$
			&	\begin{tabular}{l|l} 0	& 4 \end{tabular}
			&	\begin{tabular}{l|l} 199	& 4771 \end{tabular}
			&	\begin{tabular}{l|l} 107888	& 14886616 \end{tabular} \\ \hline
$B^{OPT}_{v,k,9}$
			&	\begin{tabular}{l|l} 1	& 2 \end{tabular}
			&	\begin{tabular}{l|l} 168	& 4164 \end{tabular}
			&	\begin{tabular}{l|l} 81901	& 8750881 \end{tabular} \\ \hline
	
$B^{OPT}_{v,k}$	
			&	\begin{tabular}{l|l} 0	& 3 \end{tabular}
			&	\begin{tabular}{l|l} 164	& 3537 \end{tabular}
			&	\begin{tabular}{l|l} 58638 & 1063939 \end{tabular} \\ \hline

\multicolumn{4}{|c|}{Rockyou} \\
\hline
Adversary type	&	Hacker	& 	Criminal	& Nation-state \Warning \\ \hline

$v/k$ ratio	&	\begin{tabular}{l|l} $10^2$	& $10^3$ \end{tabular}
			&	\begin{tabular}{l|l} $10^4$	& $10^5$ \end{tabular}
			&	\begin{tabular}{l|l} $10^6$	& $10^7$ \end{tabular} \\ \hline
			
$B^{OPT}_{v,k,5}$
			&	\begin{tabular}{l|l} 1	& 121 \end{tabular}
			&	\begin{tabular}{l|l} 1928 & 259169 \end{tabular}
			&	\begin{tabular}{l|l} 259169	& 259169 \end{tabular} \\ \hline
$B^{OPT}_{v,k,6}$
			&	\begin{tabular}{l|l} 1	& 30 \end{tabular}
			&	\begin{tabular}{l|l} 981	& 20963 \end{tabular}
			&	\begin{tabular}{l|l} 1947797	& 1947797 \end{tabular} \\ \hline
$B^{OPT}_{v,k,7}$
			&	\begin{tabular}{l|l} 0	& 18 \end{tabular}
			&	\begin{tabular}{l|l} 759	& 11077 \end{tabular}
			&	\begin{tabular}{l|l} 227299	& 2506271 \end{tabular} \\ \hline
$B^{OPT}_{v,k,8}$
			&	\begin{tabular}{l|l} 0	& 15 \end{tabular}
			&	\begin{tabular}{l|l} 417	& 7919 \end{tabular}
			&	\begin{tabular}{l|l} 168816	& 2966037 \end{tabular} \\ \hline
$B^{OPT}_{v,k,9}$
			&	\begin{tabular}{l|l} 1	& 20 \end{tabular}
			&	\begin{tabular}{l|l} 366	& 6708 \end{tabular}
			&	\begin{tabular}{l|l} 2191039	& 2191039 \end{tabular} \\ \hline
	
$B^{OPT}_{v,k}$	
			&	\begin{tabular}{l|l} 0	& 7 \end{tabular}
			&	\begin{tabular}{l|l} 320	& 6327 \end{tabular}
			&	\begin{tabular}{l|l} 114760 & 14344391 \end{tabular} \\ \hline

\end{tabular}	
}	

\end{table}

\section{Performance Comparison AES-GCM vs AES-CBC}
We ran our own measurements to compare the performance between CBC and GCM. The results are shown in 
 Figure \ref{CBCvsGCM}. First, we are producing a set of randomized bits to encrypt using the SecureRandom function in Java and storing it in a 64 byte buffer.   Next, we then take the 64 byte buffer and encrypt it 100,000 times on a 2.9 GHz Intel Core i7 with 16 GB of LPDDR3 memory.   We repeat the test 100 times and take the median score for the duration of time to hash or encrypt the randomized data with or without padding. The results are consistent  with  prior findings that GCM is substantially faster than CBC mode due to  native support for GCM at the processor level. We remark that the use of the padding flag in TLS has minimal effect on the impressive performance of GCM.

\begin{figure}
 \includegraphics[width=90mm,scale=0.5]{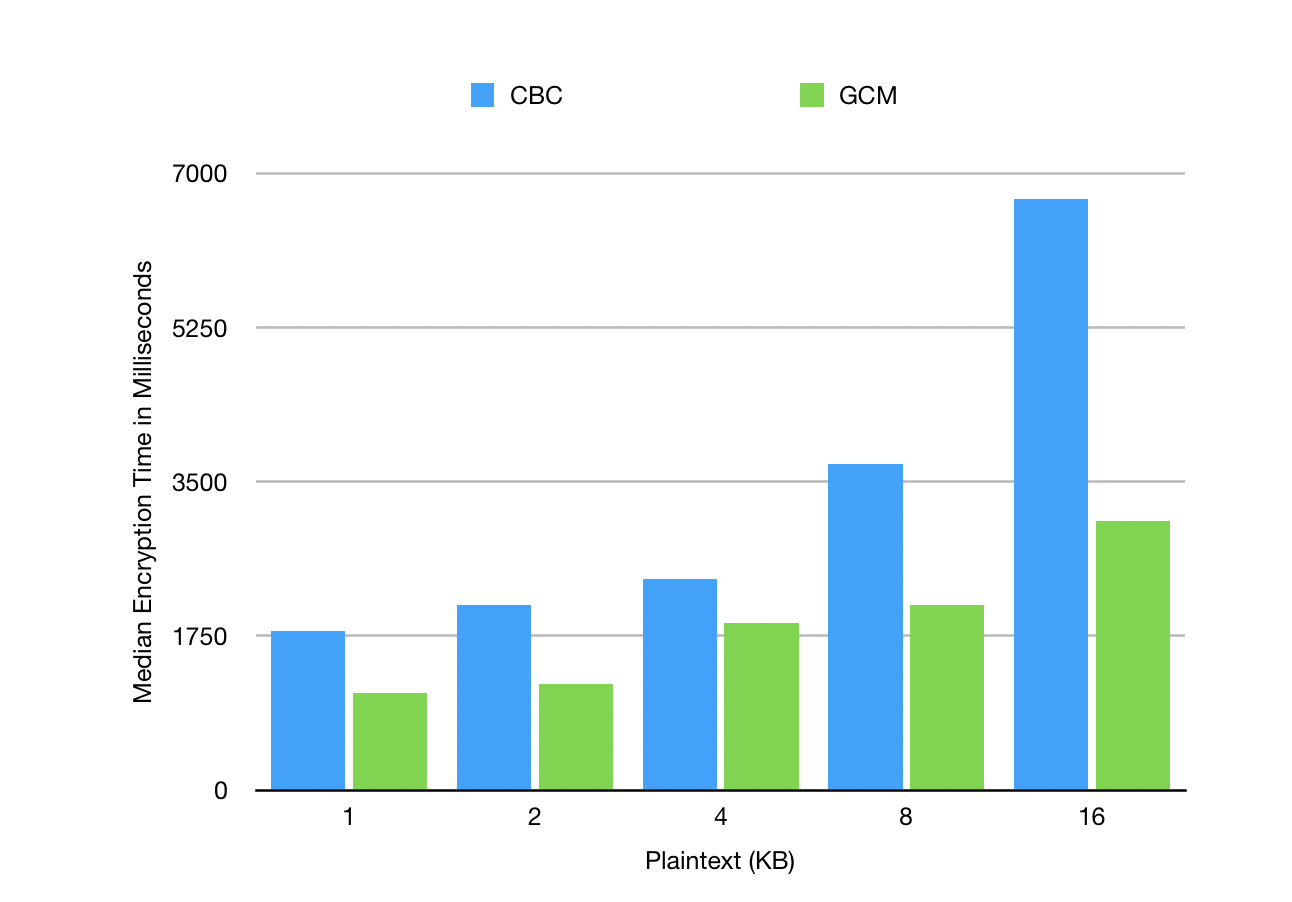}
 \caption{Comparison of performance for CBC versus GCM}
 \label{CBCvsGCM}
\end{figure}

\section{Additional Detail for Observational Studies}
\begin{figure}
\centering
\begin{tikzpicture}

\begin{axis}[
	xlabel={Password length},
	ylabel={Packet payload length},
	legend entries={IRC \& Gmail, JP Morgan Chase},
	legend pos= north west,
]

\addplot[
	black,
	samples=20,
	mark=o,
	domain=0:20,
]
{x};
\addplot[
	black,
	dashed,
	samples=20,
	mark=*,
	domain=0:20,
]
{2*x};

\end{axis}
\end{tikzpicture}
\caption{Relationship between password length and packet payload length}\label{fig:packetlength}
\end{figure}
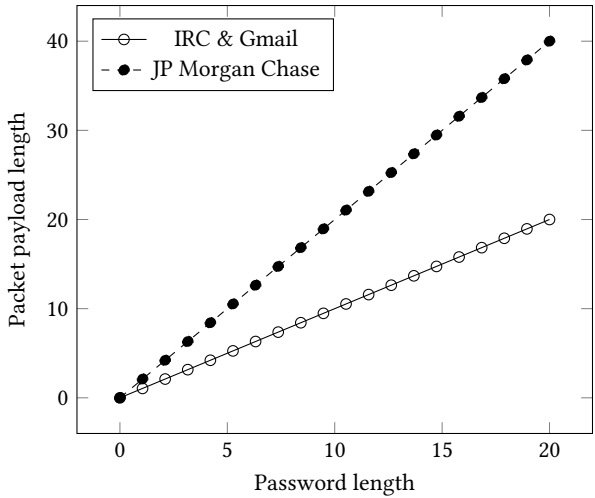

	The general approach for these studies was to observe network traffic for each service or function studied with transport security protocols: QUIC, TLS version 1.2 and TLS version 1.3 (draft).  The first study looked at Google's QUIC protocol in the Cisco ASA5506 enabled Virtual Private Network (VPN) while the second study observed QUIC during the sign-in process for Google Mail. The third and final observational study examined TLS version 1.2 during the sign in process for online banking at JP Morgan Chase. As depicted in Figure~\ref{fig:packetlength}, all three studies demonstrated an obvious pattern between encrypted traffic and the plaintext length of passwords; in particular, we observed a clear linear relationship between the password lengths and the length of the encrypted payloads in the respective packets.

\subsection*{Observational Study 1: QUIC in Cisco Virtual Private Network}
	This observational study draws inferences from a sample of network packets captured from a Cisco ASA 5506 VPN stream. This study was a necessary first step in identifying vulnerabilities in this VPN for future research.   This observational study did not control confounding variables, because this would make it difficult to understand the normal behavior of the encrypted VPN stream.

\subsubsection*{Phenomena and Variables}
	This study investigates phenomena of information leakage from encrypted traffic.  Specifically, this study collected and analyzed network traffic produced from the Cisco ASA 5506 appliance, one of the most popular VPN devices used for connecting organizations around the world.  This observational study tried to answer the following research question: what information, if any, can be derive from an encrypted Cisco VPN stream by observing the source and destination Internet Protocol (IP) address, the protocol used, the encrypted IP packet size, and the timestamps for when packets are sent.  Within this research question, we measured four (4) variables: what communication occurred, when did the communication take place, where did the communication take place, and who is involved in the communication.

\subsubsection*{Measures}
	To measure these variables, this study observed the packet attributes.  A comparison of packet attributes for encrypted and corresponding unencrypted traffic was conducted to identify any ordinal patterns among the packet sequence, source Internet Protocol address, destination Internet Protocol address, protocol, and packet length.   These network attributes were observed using packet length in bytes, the IP Addresses were in IP version 4 format, and the communication protocol was translated from Hexadecimal automatically from WireShark.  These attributes and these measures served to help us comprehend the network traffic occurring during a specific observation.

\subsubsection*{Sample}
	For each observation, the data was collected three times and checked for consistency among the collections to ensure the data was valid.  If the three collections were consistent, then we assumed that the network traffic was accurately collected for a particular network function.

\subsubsection*{Procedures}
This study used a certified Cisco network services company to set up the Cisco VPN network to ensure the lab configuration was correct.  The Cisco ASA VPN device was configured to use TLS version 1.2 to connect AnyConnect clients for remote access to the internal network. This study also utilized a collection platform that allowed simultaneous collection on the internal and external interfaces for the Cisco ASA 5506 device.  This approach allowed for easy observations to compare unencrypted traffic inside the VPN with the corresponding encrypted traffic immediately outside the VPN.  For the internal interface, this study used a 192.168.1.X C class network.  For the external interface, this study attached the Cisco ASA 5506 to a public IP Address in the C class range.  For VPN clients, this study created a specific 192.168.2.X C class pool.  This allowed the data to be distinguished between VPN clients and internal devices.  This setup allowed for easily correlation between unencrypted traffic in the 192.168.X.X range to corresponding encrypted traffic from any Internet routable IP address.  Next, the internal devices, the Internet connection, and Cisco ASA 5506 were all connected to a network hub.  The hub forwarded all network traffic to each device, and therefore, the hub allowed WireShark to sniff all internal and external communication intended to transit the VPN network.
	
	Next, this study observed common categories of Internet communication.  To represent more spontaneous types communication, this study observed messages between two persons communicating over the common Instant Relay Chat (IRC) service and to simulate more continuous types of communication, we observed file transfers using the common service File Transfer Protocol (FTP).  Next, we measured four independent variables: what, when, where, who.  To ensure accuracy, the data was collected three times and checked for consistency among the collections to ensure the data was valid. 
	 
	 For what, the data compared packet attributes for encrypted and corresponding unencrypted traffic to identify any ordinal patterns among the packet sequence, source Internet Protocol address, destination Internet Protocol address, protocol, and packet length.  For when, the data measured the interval between the first and last seen occurrence of the suspected target source and destination Internet Protocol addresses.  For where, this study used trace route and the WHOIS service to observe the source and destination Internet Protocol address to identify the city locations.  For who, this study used the WHOIS service to expose the identities of the two parties communicating. 
	
	We then reran our observational study using WireShark and allowed simultaneous collection on the internal and external interfaces for Google's Chrome web browser to better study QUIC.  Again, this approach allowed for easy observations to compare unencrypted traffic before transmission with the corresponding encrypted traffic after transmission.  A comparison of packet attributes for encrypted and corresponding unencrypted traffic was conducted to identify any ordinal patterns among the packet sequence, source Internet Protocol address, destination Internet Protocol address, protocol, and packet length.   These network attributes were observed using packet length in bytes with IP Addresses in IP version 4 format; as mentioned previously, the communication protocol was translated from Hexadecimal automatically from WireShark. During the course of our observations, it became apparent the encrypted traffic allowed for the determination of the plaintext length.   This in turn changed the focus of the research to determine how much of an advantage, if any, an attacker gained from understanding the plaintext length for passwords transmitted by QUIC, and as we detail later, the researchers used a game theoretic model to better understand how much, if any, adversarial advantage increased when QUIC leaks the plaintext length for passwords.
	
\subsubsection*{Results: QUIC Leakage in VPN}
	During collection, it was observed that the Cisco ASA 5506 switched from an encrypted TLS connection and used an experimental protocol by Google called the Quick UDP Internet Connection (QUIC).  QUIC was made to improve performance.  The QUIC protocol supports a set of multiplexed connections over UDP, and was designed to provide security protection equivalent to TLS/SSL, along with reduced connection and transport latency.  The use of QUIC is absent in the Cisco documentation.  To summarize, the identified ordinal patterns allowed observers to successfully determine the types of communication being used.  Specifically, it is possible to identify when a user browses a website or when a file is transferred.  It is also possible to positively identify when an IRC client connects and disconnects from an IRC server, distinguish when online chatting begins and ends, and even derive the total length of individual IRC chat messages. Below we briefly describe our discoveries for each of our cases:

Web Traffic:
	After analysis of the collected Web traffic, an ordinal pattern emerged. Packets corresponding to Web traffic used the QUIC protocol and had a consistent total packet length of 1392. This ordinal pattern allows an eavesdropper to create automatic signatures to determine if Web services are being used.
	
FTP Traffic:
	Here too our analysis of the FTP network traffic revealed an ordinal pattern. FTP traffic used the QUIC protocol and had a consistent packet length of 1463. As was the case with web traffic, this ordinal pattern allows an eavesdropper to create automatic signatures to determine if FTP services are being used in the VPN.

IRC Connect Traffic:
	 In our observation of IRC connection traffic, we saw that the first two packets sent during the IRC client connect have a packet length of 119 and use the TLSv1.2 protocol. The remainder of the packets that were sent during the IRC client connect are sent from the server, using the QUIC protocol, and have varying packet lengths. Each of these ordinal patterns remained consistent when the client connect phase was repeated.  It is important to note that an IRC connection is salient in identifying a chat session between two users.
	 
IRC Disconnect Traffic:	
The IRC client disconnect phase was similar to the client connect phase. The first two packets sent during the client disconnect phase have a packet length of 119 and use the TLSv1.2 protocol. The remainder of the packets are sent from the server, use the QUIC protocol, and have varying packet lengths. Again, each of these ordinal patterns remained consistent when collection of the client disconnect phase was repeated.   Again, it was important to identify an IRC disconnection to later identify a chat session between two separate users.
	 
IRC Chat Session:
 The first two packets sent during the communication phase had a packet length of 119 and use the TLSv1.2 protocol. The remainder of the packets had varying lengths, all of which were less than 300 bytes. From these initial observations, the QUIC packet sizes for IRC communication appeared to vary based upon the length of the IRC message being sent. To investigate these observations further, additional network traffic was collected. IRC messages, composed of English characters, of character length one to 100 were sent in half-second intervals while network traffic was collected. The network setup for this research allowed the unencrypted IRC packets to be collected from the internal interface of the Cisco VPN.

It is important to stress the increased risk from leaking -- via the VNP traffic -- fine-grained information about the plaintext; for example, one can easily envision a situation where the attacker can uniquely identify the two parties in an IRC chat over a VPN that otherwise would not have been possible if the leakage were less fine-grained.

\subsection*{Observational Study 2: QUIC in Google's E-mail or Google Mail}
	In this observational study, we sought to observe if QUIC specifically leaks the password length when signing into Google Mail. We again employed a collection platform using WireShark and allowed simultaneous collection on the internal and external interfaces to easily compare the encrypted and unencrypted traffic. During the course of our observations, it became apparent the encrypted traffic allowed for the determination of the plaintext length.
	
\subsubsection*{Phenomena and Variables, Measures, Sample}
The phenomena, variables, measures, and sample are the same as in the Observational Study 1.

\subsubsection*{Procedures}
The procedures changed slightly for this observational study.  In this study, we used the Chrome web browser, because by default, QUIC is enabled with Chrome.  We then used our automated process to login to Google Mail three times increasing the password length by one character every ten seconds.   We had a ten second delay between passwords entered as this helped introduce a pronounced signature when reviewing QUIC network traffic in WireShark.   Google Mail uses reCAPTCHA to slow the process down, but we were able to manually enter the CAPTCHA challenge within the ten second time delay built into our observational study.   

\subsubsection*{Results: Password Length Leakage in QUIC}
Our results suggest the password length is easily identified in traffic analysis. The encrypted packet length and the password length have a 1:1 to ratio or as the password length increases by 1 character, then the encrypted packet length increases also by 1 byte.


\subsection*{Observational Study 3: TLS version 1.2 in JP Morgan Chase Website}

	In this observational study, we observed if TLS version 1.2 specifically leaks the password length when signing into We .   As in the previous study, we employed a collection platform using WireShark and allowed simultaneous collection on the internal and external interfaces to easily compare the encrypted and unencrypted traffic. During the course of these observations, it again became apparent the encrypted traffic allowed for the determination of the plaintext length. To ensure consistency, we reviewed that TLS version 1.2 used AES-GCM every time we signed into JP Morgan Chase from our web browser.

\subsubsection*{Phenomena and Variables, Measures, Sample}
The phenomena, variables, measures, and sample are the same as in the Observational Study 1 and 2.

\subsubsection*{Procedures}
The procedures changed slightly for this observational study.  In this study, we used Apple's Safari web browser to observe TLS version 1.2 traffic.  This helped us understand that any leaking of information was not exclusive to either the web browser or transport security protocol.  We then used our automated bot to login to JP Morgan Chase's online banking website three times increasing the password length by one character every ten seconds.   We had a ten second delay between passwords entered, because this helped introduce a pronounced signature when reviewing TLS version 1.2 network traffic in WireShark.   It should be noted that JP Morgan Chase returned a lock out message, but in reality, it did not lock us out.   We are able to submit a processed query to the server each time without being blocked.   As soon as we signed in with a legitimate account, it did not require multi-factor authentication.  

\subsubsection*{Results: Password Length Leakage in QUIC}
Our results suggest the password length is easily identified in traffic analysis. The encrypted packet length and the password length have a 1:1 to ratio or as the password length increases by 1 character, then the encrypted packet length increases also by 1 byte.


\section{Differential Privacy Overview}
Differential privacy~\cite{dwork2006calibrating} is a rigorous mathematical formulation of privacy which has emerged as the {\em de facto} standard. Intuitively, differential privacy captures the right of any individual user to withhold his or her personal data from a study. Informally,  differential privacy ensures that any inference that can be made could have been made whether or not a particular individual's data was excluded. Differential privacy does not necessarily rule out publishing accurate aggregate statistics e.g., how many hospital patients smoke, but would necessarily rule out (accurate) targeted statistics e.g., how many patients named John Doe live in city X and have cancer?  

\paragraph{Password Frequency Lists} In our setting a password dataset $D$ consists of a list tuples $(u,pwd_{u})$ where $pwd_u$ is the password selected by user $u$. Given a dataset $D$ the password frequency consists of non-negative list of integers $f_1 \geq f_2 \geq \ldots \geq 0$ where $f_i$ denotes the number of users in a dataset who selected the $i$\textsuperscript{th} most popular password and $N= \sum f_i$ denotes the total number of users in the dataset We also use $f_1^\ell \geq f_2^\ell \geq \ldots \geq 0$ to denote the frequency list for length $\ell$ passwords, and $N_\ell = \sum f_i^\ell$ to denote the total number of users who selected a password of length $\ell$ --- $f_i^\ell$ denotes the number of users who selected $i$\textsuperscript{th} most popular password with length $\ell$. 

As an example consider a toy scenario in which $12$ users create LinkedIn passwords. Suppose that $f_1= 5$ users select the password ``123456,'' $f_2=3$ users select the password ``password'' and $f_3 = 2$ users select the password ``abc123'' and $f_4 = 2$ users select the password ``letmein.'' In this case the overall frequency list would be $(f_1,f_2,f_3,f_4) = (5,3,2,2)$ with $N=12$, but the frequency list for length $\ell = 6$ passwords would be $(f_1^6,f_2^6)=(5,2)$ with $N_6 = 7$ total passwords of length $6$. 

We remark that a password frequency list is de-anonymized in the sense that the actual usernames and passwords have been omitted. However, we stress that releasing a password frequency list without noise does not satisfy the stringent requirements of differential privacy. 

\paragraph{Security Risks}  We first discuss the potential risk of releasing password frequency lists without any noise. In the example above that the attacker happened to know the passwords for eleven of these users. If only four of these eleven users selected the password ``123456'' then the attacker who obtains the exact frequency list $(f_1,f_2,f_3,f_4) = (5,3,2,2)$ above would be able to infer the password of the remaining user. While this scenario may seem a bit far-fetched there are many examples of supposedly anonymized datasets that were later de-anonymized when the attacker has some background knowledge about the data e.g., \cite{SP:NarShm09}.

\paragraph{Formal Definition of Differential Privacy for Password Frequency Lists}
A mechanism $\mathcal{A}$ which takes as input a frequency list $f_1,f_2,\ldots$ with $f_i \geq f_{i+1}$ and outputs a noisy frequency list is $(\epsilon,\delta)$-differentially private if for all subsets $S$ of output frequency lists we have

\[\Pr\left[\mathcal{A}\left(f\right) \in S \right] \leq e^{\epsilon} \Pr\left[\mathcal{A}\left(f'\right) \in S \right] + \delta \ , \]

Blocki et al.~\cite{NDSS:BloDatBon16} previously developed an $(\epsilon,\delta)$-differentially private algorithm to publish (noisy) password frequency lists. While their algorithm does add some noise to the data they proved that the L1 error was small i.e. \[ \sum_{i} \left| f_i - \tilde{f}_i\right| \leq O\left( \frac{\sqrt{N}+\ln\left(\frac{1}{\delta} \right)}{\epsilon} \right) \ , \] where $f$ is the original frequency list and $\tilde{f}$ is the noisy frequency list output by their mechanism. Empirical analysis shows that the L1 error is small in practice~\cite{NDSS:BloDatBon16,blockidifferentially}. 

\paragraph{Discussion} Differential privacy has several attractive features. It provides rigorous guarantees even if the attacker already has significant background knowledge about one or more user's in the dataset. Another nice feature of differential privacy is that it is preserved under post-processing~\cite{dpbook}. Intuitively, we can take this mean that for any attack that an adversary might mount against a particular individual user $u$ the probability that the attack succeeds (e.g., guesses the user's password) would essentially be the same even if we had excluded the user's password $(u,pwd_u)$ from the original dataset before publishing our noisy frequency list $\tilde{f}$. This guarantee holds even if the adversary already knows the passwords for all $n-1$ other user's. In this sense $\tilde{f}$ will not help the attacker to guess any particular individual's password. We refer an interested reader to \cite{NDSS:BloDatBon16} for more discussion of differential privacy as it applies specifically to password frequency lists, and we refer an interested reader to the excellent textbook of Roth and Dwork~\cite{dpbook} for more discussion of differential privacy in general. 


\end{document}